\documentclass[12pt,draftcls,onecolumn]{ieeecolor}

\usepackage{epsfig} 
\usepackage{generic}
\usepackage{setspace}
\usepackage{amsmath} 
\usepackage{amssymb}  
\usepackage{amsfonts}
\usepackage{color}
\usepackage{booktabs}
\usepackage{comment}
\usepackage{placeins}

\newtheorem{definition}{Definition}
\newtheorem{remark}{Remark}
\newtheorem{problem}{Problem}
\newtheorem{theorem}{Theorem}

\newtheorem{lemma}{Lemma}
\newtheorem{example}{Example}

\DeclareMathOperator{\diag}{diag}

\newcommand{\field}[1]{\mathbb{#1}}

\newcommand{\N}{\field{N}}

\newcommand{\vvss}{``}
\newcommand{\vvdd}{''}
\newcommand{\vv}[1]{\vvss #1\vvdd}

\newcommand{\Registered}{\ensuremath{^{\hbox{\fontsize{5pt}{5pt}\selectfont%
      \textregistered}}}}

\title{\LARGE \bf {Novel Conditions for the Finite-Region
Stability of 2D-Systems with
Application to Iterative Learning
Control}}

\author{
Chao Liang\thanks{C. Liang is with the School of Automation, Nanjing University of Science and Technology, Nanjing, Jiangsu, China.} ,
Carlo Cosentino$^\ast$, Alessio Merola\thanks{C. Cosentino and A. Merola are with the School of Computer and Biomedical Engineering, Dipartimento di Medicina Sperimentale e Clinica, Universit\`a degli Studi Magna Gr{\ae}cia di Catanzaro, Campus di Germaneto \vv{Salvatore Venuta}, 88100 Catanzaro, Italy.} , Maria Romano and Francesco Amato\thanks{M. Romano and F. Amato are with the Dipartimento di Ingegneria Elettrica e delle Tecnologie dell'Informazione, Universit\`a degli Studi di Napoli Federico II, Via Claudio 21, 80125 Napoli, Italy.\\$^\ast$Corresponding author (\texttt{carlo.cosentino@unicz.it}).}
}

\begin{document}

\maketitle
\thispagestyle{empty}
\pagestyle{empty}


\begin{abstract}
 Some recent papers have extended the concept of finite-time stability (FTS) to the context of 2D linear systems, where it has been referred to as finite-region stability (FRS). FRS methodologies make even more sense than the classical FTS approach developed for 1D-systems, since, typically, at least one of  the state variables of 2D-systems is a  space coordinate, rather than a time variable. Since space coordinates clearly belong to finite intervals, FRS techniques are much more effective than the classical Lyapunov approach, which looks to the asymptotic behavior of the system over an infinite interval. To this regard, the novel contribution of this paper goes in several directions. First, we provide a novel sufficient condition for the FRS of linear time-varying (LTV) discrete-time 2D-systems, which turns out to be less conservative than those ones provided in the existing literature. Then, an interesting application of FRS to the context of iterative learning control (ILC) is investigated, by exploiting the previously developed theory. In particular, a new procedure is proposed so that the tracking errors of the ILC law converges within the desired bound in a finite number  of iterations. Finally, a sufficient condition to solve the finite-region stabilization problem is proposed. All the results provided in the paper lead to optimization problems constrained by linear matrix inequalities (LMIs), that can be solved via widely available software. Numerical examples illustrate and validate the effectiveness of the proposed technique.


\end{abstract}

\section{Introduction}\label{section: Introduction}

The concept of finite-time stability (FTS)~\cite{FTS_book}  has been introduced to characterize the behavior of a given system over a bounded time interval; in this time interval the trajectories of the system should remain inside a given domain (\textit{trajectory domain}), whenever the initial conditions belong to another given domain (\textit{initial domain}). Hence, as it was originally introduced in~\cite{Lebedev_jamm_1954a,Dorato:61}, FTS does not consider what happens asymptotically, as it is the case for Lyapunov stability, but it rather considers only \textit{short} time intervals.

Following the publication of~\cite{Amato01,Amato05}, where the authors considered the FTS of linear time-invariant (LTI) systems, ending up with convex conditions based on the solution of a set of linear matrix inequalities (LMIs), a great effort has been spent to  extend such approach to linear time-varying (LTV)  systems~\cite{Amato_survey_06,Tarb09,Amato10,Amato_aut_2010,Amato_aut_2017}. LTV systems play an important role in several engineering problems, such as  chemical processes, spacecraft control and power systems \cite{Galiaskarov2017,Bhat2005,Belikov2018}. 
Although, for the sake of simplicity, in many applications the time-varying nature of the system is ignored, with the increasing requirements of modern control systems,  the effect of time-varying phenomena on the performance of a given system can become significant, and therefore the time-varying characteristics of the system itself must be taken into consideration.

In order to deal with time-varying systems, an approach based on time-varying quadratic Lyapunov functions has been exploited; such approach  leads to conditions for analysis and design in terms of feasibility problems involving differential LMIs (in the continuous-time case) or difference LMIs (DLMIs) (in the discrete-time context); it is worth noting that the use of time-varying Lyapunov functions allows to derive non-conservative, i.e. necessary and sufficient conditions for FTS \cite{Amato_aut_2013}, which therefore improve the approach of the former papers \cite{Amato01,Amato05}, even when LTI systems are dealt with (see also \cite{FTS_book}, Chapter 2, and the bibliography therein). Finally,
the FTS problem has been also investigated in the context of uncertain systems \cite{Amato_ijc_2011b}, nonlinear systems \cite{Amato_tac_NLQS}, hybrid systems~\cite{NAHS2010,Zhao08}.

On the other hand, two-dimensional (2D) linear systems \cite{Roesser,Fornasini2} play an important role in many application fields, such as digital filtering,  image processing, pattern recognition, etc. (see~\cite{Kakzorek,Fornasini}). To this regard, the FTS approach has been recently extended to the class of 2D-systems,  where it has been referred to as finite-region stability (FRS) (see \cite{Hua_2018}-\nocite{Amato2017,Zhang_2017,Hua_2018_b,Gao,YANG20201,YANG202011974}\cite{LI2024101433}).

It is important to point out that the FTS methodologies extended to 2D-systems make even more sense than the FTS approach developed for 1-D systems, since the state variables of 2D-systems often depend on a pair of space coordinates, rather than on time, as it is the case, for example, of the image processing context; in some other cases one independent variable is time and the other one is a space coordinate, when distributed systems are dealt with. In these cases, the classical Lyapunov approach, which looks to the asymptotic behavior of the system over an infinite interval, is not applicable, since space coordinates  clearly belong to finite intervals.

In this context, the first contribution of our work consists of a novel sufficient condition for the FRS of 2D linear time-varying systems, obtained by exploiting the properties 
of what will be called the {\em $+$ operator}; 
also, two-dimensional time-varying quadratic Lyapunov functions are used, which allows to reduce the conservativeness of the proposed methodology with respect to the existing literature. It is worth noting that the FRS proof for 2D linear systems provided in this paper is much more challenging than both the corresponding FTS proof for 1D systems (see \cite{FTS_book}), and the proof of classical Lyapunov stability for 2D systems (see \cite{Kakzorek}); roughly speaking, this is due to the finiteness of the region in which the Lyapunov first difference has to be negative definite, when FRS is dealt with.

Starting from the analysis condition, another contribution of this paper is a methodology for the design of state feedback controllers which finite-region stabilize a given 2D-system. Both the analysis and synthesis results require the solution of feasibility problems constrained by coupled DLMIs and LMIs, that can be solved via widely available software.

Another important application that can be recast in the field of 2D-systems is the trajectory tracking problem \cite{Kurek1993}, which represents an hot research topic in control theory. Roughly speaking, the task of the  tracking problem  is to design a controller so that the controlled output follows as close as possible the desired trajectory over a given time interval.  
In this context, iterative learning control (ILC), for its great potentials in improving the performance of the tracking  controller, has been widely investigated  in recent years \cite{Yang2016,Chi2018,Shen2016}.

ILC is a kind of intelligent control approach, which can be traced back at least to forty years ago (see, among others, \cite{Arimoto,Craig,Moore,Moore_automatica,Bristow,Ahn}). The essential idea of ILC is that the control signal is iteratively updated,  on the basis of the information obtained from previous iterations, to track the desired trajectory with more accuracy. Since the Seventies, ILC  has been applied to many contexts, such as industrial robots \cite{Norrlof2002}, multi-agent systems  \cite{Jin2016} and permanent-magnet spherical actuators \cite{Zhang2016}.

As said, in \cite{Kurek1993} it has been shown that the  algorithm involved in the tracking control law can be recast in the framework of 2D-systems theory; in the same paper the convergence issues have been also investigated. In \cite{Fang2003,Meng2015}, the 2D analysis approach was exploited to design a suitable ILC law for systems with variable initial conditions. In these papers, the ILC approach is analyzed on the basis of a robust stability approach in the context of 2D-systems, so that the tracking error converges to zero as the number of iterations approaches  infinity.  

As far as we know, the analysis and design of ILC laws over finite intervals has been rarely investigated. Nevertheless, in some cases of practical engineering applications, due to the increased requirements for tracking performance, it is required that, in a finite number of iterations, the iterative learning controller  guarantees  the convergence within the desired bound of the corresponding tracking error. Motivated by this consideration, FRS of 2D-systems, differently from asymptotic stability,  is a more practical concept to be exploited to analyze the trajectory tracking problem in such cases.

To this regard, another contribution of this paper shows how a ILC problem, defined over a finite interval of time, can be reformulated as a FRS problem for a particular 2D-system. Exploiting the sufficient conditions  for FRS developed in the first part of the paper, a new ILC algorithm is proposed, which converges in a finite  number of iterations, as well as the associated tracking error. The convergence of the proposed algorithm is investigated and sufficient conditions for convergence, based on feasibility problems constrained by LMIs, are proposed.

The paper is organized as follows. In~Section \ref{sec:2}, we precisely state the problem we deal with, and some preliminary results are given. In Section~\ref{sec:3} a novel sufficient condition for the FRS of 2D linear systems is provided, together with a numerical example illustrating the proposed approach; moreover we show that our result is less conservative with respect to the existing literature.  Then, the application of the FRS approach to solve the ILC problem over a finite-interval is illustrated, together with an example to show the effectiveness of the proposed methodology. In Section~\ref{sec:2_b} the finite-region stabilization problem is investigated.
 Finally, in Section~\ref{sec:conclusions}, some concluding remarks are given.

\noindent
{\bf Notation.} The symbol $\N_0$ denotes the set of integer numbers, i.e. $\{0,1,2,\dots\}$. Given the sets $S$ and $V$, with $V\subseteq S$, $S/V$ denotes the set composed of the elements of $S$ which are not elements of $V$. The matrices, if not explicitly stated, are always assumed to have compatible dimensions.
$I$ denotes the identity matrix. 
$A^T$ stands for the matrix transpose of $A$. $P>0$ ($P\geq 0$) means that $P$ is symmetric, $P=P^T$, and positive definite (semidefinite).
The symbol  $\diag(A,B)$ denotes the block-diagonal matrix having the matrices $A$ and $B$ on the diagonal.

\section{Preliminary Definitions and Results} \label{sec:2}
In this section, we recall the extension of the FTS concepts to 2D-systems and how the ILC design can be framed in the 2D-system context; moreover, some
technical tools, needed for the development of our machinery, will be provided. 

\subsection{Finite-Region Stability of 2D-Systems}
Let us consider a  bidimensional grid, in which the pair of indices~$(h,k)\in\N_0\times\N_0$, 
univocally individuates a point of the grid itself. Unlike 1-D systems, the indices $h$ and $k$, do not necessarily represent time variables; if, for instance, we think to an image, then the pair~$(h,k)$ is associated to a specific pixel of the image. The coordinates~$h$ and~$k$ are defined as the vertical and the horizontal coordinate, respectively.

In the following,  we consider a discrete-time linear 2D-system in Roesser form, namely (see \cite{Roesser})
\begin{equation}\label{2Dsys}
    \begin{pmatrix}
x_1(h+1,k)\\x_2(h,k+1)
\end{pmatrix}=\begin{pmatrix} A_{11}(h,k) & A_{12}(h,k) \\
A_{21}(h,k) & A_{22}(h,k)
\end{pmatrix} \begin{pmatrix}
x_1(h,k)\\x_2(h,k)
\end{pmatrix}\,, 
\end{equation}
where $A_{ii}(h,k)$, $i=1,2$, are square 2D matrix-valued sequences, and~$A_{12}(h,k)$, $A_{21}(h,k)$ are of compatible dimensions; the 2D vector-valued sequences~$x_1(h,k)$ and~$x_2(h,k)$ represent the vertical and the horizontal state of the system;  we define the compact vector $x(h,k):=\begin{pmatrix} x_1^T(h,k) & x_2^T(h,k)\end{pmatrix}^T$,

For a comprehensive treatment of 2D-systems, the interested reader is referred to \cite{Kakzorek}.


Given the finite set~$\mathcal N_1\times \mathcal N_2:=\{0,1,\dots, N_1\}\times \{0,1,\dots,N_2\}\subset \N_0\times\N_0$, 
we recall the definition of FRS, which extends the FTS concept to 2D-systems~\cite{Hua_2018,Amato2017,Zhang_2017,Zhang_2016}.
\begin{definition}\label{def_FSS}
Given the finite set $\mathcal N_1\times \mathcal N_2$, the symmetric positive definite 2D matrix-valued sequence $R(h,k):=\diag\left(R_1(k),R_2(h)\right)$, and the symmetric positive definite 2D matrix-valued sequence $\Gamma(h,k):=\diag\left(\Gamma_1(h,k),\Gamma_2(h,k)\right)$, with $\Gamma(0,0)<R(0,0)$, $\Gamma_1(0,k)<R_1(k)$, $k\in\mathcal N_2$, $\Gamma_2(h,0)<R_2(h)$, $h\in\mathcal N_1$, system \eqref{2Dsys} is said to be finite-region stable with respect to $(\mathcal N_1 \times \mathcal N_2,R(h,k),\Gamma(h,k))$, if, for all $(h,k) \in \mathcal N_1\times \mathcal N_2$,
\begin{equation}\label{FTSdef}
   \sum\limits_{k=0}^{{{N}_{2}}}{x_{1}^{T}(0,k)}R_1(k){{x}_{1}}(0,k)+\sum\limits_{h=0}^{{{N}_{1}}}{x_{2}^{T}(h,0)}R_2(h){{x}_{2}}(h,0)\le 1,
\end{equation}
implies
\begin{equation} \label{FTS_cond_state}
    x^T(h,k)\Gamma(h,k)x(h,k)<1, \quad\forall(h,k) \in\mathcal N_1 \times \mathcal N_2\,.
\end{equation}
\hfill$\Diamond$
\end{definition}

Definition \ref{def_FSS} generalizes, in a natural way, the classical definition of FTS (see~\cite{FTS_book}, Ch.~5) to linear discrete-time 2D-systems. Indeed, condition \eqref{FTSdef} is consistent with the fact that, in order to compute the solution of a two dimensional system for all $h,k>0$, it is necessary and sufficient to know the vectors $x_1(0,k)$ and $x_2(h,0)$ 
(see \cite{Kakzorek}, Ch. 1). That is, such vectors represent the initial conditions for a 2D-system.

\subsection{The $+$  Operator} \label{plus}
For the following developments, given a 2D vector-valued sequence, we define the forward shift operator denoted by $+$, as follows
\begin{equation} \label{def+}
 {x}^{+}(h,k):=
  \begin{pmatrix}
  {{x}_{1}}(h+1,k)  \\
  {{x}_{2}}(h,k+1)  
\end{pmatrix} \,.
\end{equation}

By using \eqref{def+}, system \eqref{2Dsys} can be rewritten in the following compact form
\begin{equation} \label{2dsys_compact}
    x^{+}(h,k) = A(h,k) x(h,k)\,, 
\end{equation}
where
\begin{equation} \label{defA}
A(h,k):=\begin{pmatrix}
  A_{11}(h,k) & A_{12}(h,k) \\
A_{21}(h,k) & A_{22}(h,k)
\end{pmatrix}\,. 
\end{equation}

Now, let $x(\cdot,\cdot)$ any trajectory of system \eqref{2dsys_compact}; given a symmetric block-diagonal 2D matrix-valued sequence, partitioned compatibly with \eqref{2Dsys},
\begin{equation}\label{defP}
  P(h,k):=\left( \begin{matrix} P_1(h,k) & 0 \\
0 & P_2(h,k)
\end{matrix} \right),
\end{equation}
where $P_1(\cdot,\cdot)$ and $P_2(\cdot,\cdot)$ are positive definite matrix-valued sequences, let us consider the quadratic Lyapunov function, evaluated along the trajectory $x(\cdot,\cdot)$, $V(h,k):=x^T(h,k)P(h,k)x(h,k)$.

If we define, consistently with \eqref{2Dsys},
\begin{equation}\label{defP+}
  P^{+}(h,k):=\left( \begin{matrix} P_1(h+1,k) & 0 \\
0 & P_2(h,k+1)
\end{matrix} \right)\,, 
\end{equation}
we can compute the forward shift of $V$ along the trajectory $x(\cdot,\cdot)$,  induced by the $+$ operator, as follows
\begin{align} \label{def_V_compact}
    V^+(h,k)
     &= {x^+}^T(h,k) P^{+}(h,k) x^+(h,k)\vert_{x^+(h,k)=A(h,k)x(h,k)} \nonumber \\
     &= x^T(h,k) A^T(h,k) P^{+}(h,k) A(h,k)x(h,k) \nonumber \\
    &\hspace{-1cm} =x_1^T(h+1,k)P_1(h+1,k)x_1(h+1,k)+x_2^T(h,k+1)P_2(h,k+1)x_2(h,k+1)\,.\nonumber \\ 
\end{align}

\subsection{ILC and 2D-Systems} \label{ILC2Dsys}
Let us consider the following 1-D discrete-time linear time-varying system
\begin{subequations}\label{sistema}
\begin{align}
 \label{sistema_a}   x(t+1)&=A(t)x(t) + B(t)u(t) \,,\quad x(0)=x_0,\\
    y(t)&=C(t)x(t) \,,
\end{align}
\end{subequations}
where  $t\in \{0,1,\ldots,T\}=:\Omega$ is the discrete-time index, $A(t)$, $B(t)$, and $C(t)$ are matrix-valued sequences of compatible dimensions, ${x}(t)$, ${u}(t)$  and ${y}(t)$  are the state, input and output vectors, respectively.

The  objective is to design an iterative learning law $u(t)$, such
that the system output $y(t)$ tracks the desired output trajectory $y_r(t)$ as close as possible on the interval~$\Omega^+:=\Omega/\{0\}$. To achieve this goal we shall exploit an ILC approach.

A general ILC rule can be given as follows \cite{Kurek1993}
\begin{equation} \label{control_law}
    u(t,k+1)=u(t,k)+\Delta u(t,k)\,,
\end{equation}
where $k$ is the iteration index.

Note that, if we take into account system~\eqref{sistema} and the control update rule~\eqref{control_law}, we obtain a process which is governed by two time scales. The former is the usual time variable $t$; the latter is the iteration index $k$ of the control rule. According to \cite{Kurek1993}, system \eqref{sistema}, together with condition \eqref{control_law}, can be rewritten as a 2D Roesser system, as follows
\begin{subequations}\label{sysink}
\begin{align}
    x(t+1,k)&=A(t)x(t,k) + B(t)u(t,k)\,,\quad  x(0,k)=x_0 \\
    u(t,k+1)&=u(t,k)+\Delta u(t,k)\,,\quad u(t,0)=0\\
    y(t,k)&=C(t)x(t,k) \,.
\end{align}
\end{subequations}

Let us introduce the tracking error, as
\begin{equation}
    e(t,k)=y_r(t)-y(t,k)\,.
\end{equation}

We use the following learning law to modify  the control input
\begin{equation}    \label{update}
    \Delta u(t,k)=K(t,k)e(t+1,k)\,.
\end{equation}

In most ILC works, it is always required that the tracking error converges to zero as the iteration index $k$
approaches infinity. However, in practical engineering applications, due to the increased
requirements for tracking performance, we require that the tracking error converges, within a desired bound,
in a finite number of iterations. Therefore, as first step, we need to fix the allowable number of iterations such that the error converges within the assigned bound.

\begin{remark}\label{remark_N}
According to the requirements of practical engineering applications, we can {\em a priori} fix the maximum allowable number of iterations $N$.  For instance, if the sampling time of system \eqref{sistema} for the whole iteration experiments 
 is $T_s$, we need that the maximum number of iterations, say $N$, is such that
\[N \cdot T_w  \leq  T_s\,,\]
where $T_w$ is the estimated average time that is taken by one iteration of the algorithm which evaluates $\Delta u$.
\hfill$\blacksquare$
\end{remark}

Therefore, from now on, we shall assume that $k\in \{0,1,\dots,N\}=:\mathcal N$, where $N$ can be chosen on the basis of the considerations done in Remark \ref{remark_N}.


In the sequel we shall investigate the following problem.
\begin{problem} \label{p1}
Given a 2D symmetric positive definite matrix-valued sequence $\tilde{\Gamma }(t,k)$, defined over $\Omega \times \mathcal N$, design an iterative control law in the form \eqref{control_law}, \eqref{update}, such that
\begin{equation}\label{etN}
e^T(t,N)\tilde{\Gamma } (t,N)e(t,N) < 1\,, \forall t\in \Omega^+\,.
\end{equation}

\hfill$\blacksquare$
\end{problem}

\begin{remark} \label{preproof}
 It is worth noting that, since we are only interested in keeping the error norm below the threshold at $k=N$ (see \eqref{etN}), the matrix weight $\tilde{\Gamma }(t,k)$, for $k=0,1,\dots,N-1$, turns out to be a degree of freedom that can be exploited to speed up the error convergence at time $t$. In other words, at a given $t$, $\tilde{\Gamma }(t,k)$ can be designed in order to let the corresponding ellipsoid shrinking more and more toward the ellipsoid $\left\{e\,: e^T \tilde{\Gamma }(t,N)e < 1\right\}$, as $k$ increases, according to a desired rate of convergence. 
\hfill$\blacksquare$
\end{remark}

One way to choose the matrix weight $\tilde{\Gamma }(t,k)$, is that of generalizing the approach of the previous literature on ILC. Let us factor $\tilde{\Gamma }(t,k)$ as
\begin{equation} \label{weight_fix}
    \tilde{\Gamma }(t,k)=\Theta(t) \rho^{N-k}\,, \quad (t,k)\in\Omega^+\times \mathcal N\,,
\end{equation}
where $0<\rho<1$, and $\Theta(t)$ is a symmetric positive definite matrix sequence chosen on the basis of practical considerations involving the system dynamics.
According to \eqref{weight_fix}, we have, for a given $t\in\Omega^+$,
\begin{align} \label{deflation}
    \rho^N \Theta(t) & < \rho^{N-1} \Theta(t) \nonumber \\
     & < \cdots \nonumber \\
     & < \rho^{N-k} \Theta(t) \nonumber \\
     & < \cdots \nonumber \\
     & < \Theta (t)\,.
\end{align}

According to \eqref{weight_fix}-\eqref{deflation}, we have that, for a given $t\in\Omega^+$,
\begin{align}
\left\{e\,: e^T\tilde{\Gamma }(t,0)e < 1\right\} & \supset \left\{e\,: e^T \tilde{\Gamma }(t,1)e < 1\right\} \nonumber \\
     & \supset \cdots \nonumber \\
     & \supset  \left\{e\,: e^T \tilde{\Gamma }(t,k)e < 1\right\} \nonumber \\
     & \supset \cdots \nonumber \\
     & \supset \left\{e\,: e^T \tilde{\Gamma }(t,N)e < 1\right\}\,,
\end{align}
i.e, following the idea of Remark \ref{preproof}, we constraint, at each time $t$, the error to stay inside $N$ ellipsoids of decreasing size as $k$ increases, until the terminal ellipsoid  is reached.
\hfill$\blacksquare$



\bigskip
\subsection{ILC and FRS} \label{ILCFRS}
In order to recast system \eqref{sysink} in a format which is suitable of a FRS interpretation, let us rewrite \eqref{sysink}, according to \cite{Kurek1993}, in the incremental state-error form, by letting $\eta(t,k)=x(t-1,k+1)-x(t-1,k)$, as follows
\begin{subequations}\label{Roesser}
\begin{align}
    \begin{pmatrix}
\eta(t+1,k) \\ e(t,k+1)
\end{pmatrix}
&=\begin{pmatrix}A(t-1)&B(t-1)K(t-1,k)\\-C(t)A(t-1)&I-C(t)B(t-1)K(t-1,k)\end{pmatrix}\begin{pmatrix}
\eta(t,k)\\e(t,k)
\end{pmatrix} \nonumber\\
&=: A_{\text{ILC}}(t,k)\begin{pmatrix}
\eta(t,k)\\e(t,k)
\end{pmatrix},\\
\begin{pmatrix}
\eta(1,k) \\
e(t,0)
\end{pmatrix} &= \begin{pmatrix}
x(0,k+1)-x(0,k) \\
y_r(t)-y(t,0)
\end{pmatrix}=\begin{pmatrix}
0\\
y_r(t)-C(t)x(t,0)
\end{pmatrix}\,,
\end{align}
\end{subequations}
where $(t,k)\in \Omega^+\times\mathcal N$.

Note that $e(t,0)$ depends on $x(t,0)$, that can be computed as the zero-input solution of equation \eqref{sistema_a};
therefore $e(t,0)$ can be considered a problem data.

Let $\tilde {R}(t)$ any symmetric positive definite matrix-valued sequence such that, for all $t\in\Omega^+$,
\begin{equation} \label{condition_R}
   \sum\limits_{t\in\Omega^+} e^T(t,0)\tilde {R}(t)e(t,0)\leq 1\,, \quad \tilde R(t)> \tilde \Gamma(t,0)\,,
\end{equation}
then the following technical result puts in connection FRS and ILC.
\begin{lemma} \label{FTSforILC}
Problem \ref{p1} is solvable if
there exist scalars $\epsilon_1>\epsilon_2 >0$, such that system \eqref{Roesser} is finite-region stable wrt $\left(\Omega^+\times \mathcal N,\right.$
$\diag(\epsilon_1 I,\tilde {R}(t)),$
$\left.\diag(\epsilon_2 I, \tilde{\Gamma}(t,k)\right)$.
\end{lemma}
Before proving the lemma, some considerations are in order.
\begin{remark}
The choice of the structure of the initial state weight in the form $\diag(\epsilon_1 I,\tilde {R}(t))$, where $\epsilon_1$ is left free for optimization purposes, follows from the fact that we are only interested in the behaviour of the error $e(t,k)$, i.e. the second component of the state of system \eqref{Roesser}; the same consideration holds for the trajectory weight $\diag(\epsilon_2 I, \tilde{\Gamma}(t,k))$.
\hfill$\blacksquare$
\end{remark}

\begin{proof}

The assumption of the lemma guarantees that there exist $\epsilon_1>\epsilon_2>0$, such that, for all pair  $(t,k)\in\Omega^+\times \mathcal N$, the condition
\begin{equation}
   \sum\limits_{k\in\mathcal N} \epsilon_1 \eta^T(1,k)\eta(1,k)+\sum\limits_{t\in\Omega^+} e^T(t,0)\tilde {R}(t)\, e(t,0) \leq 1,
\end{equation}
implies
\begin{equation}
    \epsilon_2 \eta^T(t,k)\eta(t,k)+e^T(t,k)\tilde{\Gamma} (t,k) e(t,k) <1\,. 
\end{equation}

The above condition holds, in particular, by choosing an identically zero initial condition for $\eta(t,k)$, i.e. $\eta(1,k)=0$, $k\in\mathcal N$;
therefore we have the following chain of implications
\begin{align} \label{et0}
    \sum\limits_{t\in\Omega^+} e^T(t,0)\tilde {R}(t) \, e(t,0) \leq 1 & \Rightarrow \epsilon_2 \eta^T(t,k)\eta(t,k)+e^T(t,k)\tilde{\Gamma}(t,k) e(t,k) <1 \,,\nonumber \\ & \hspace{4.8cm} \forall (t,k)\in\Omega^+\times \mathcal N \nonumber \\
    & \Rightarrow e^T(t,k)\tilde{\Gamma}(t,k) e(t,k) <1 \,, \quad\forall (t,k)\in\Omega^+\times \mathcal N \nonumber \\
    & \Rightarrow e^T(t,N)\tilde{\Gamma}(t,N) e(t,N) <1 \,, \forall t \in \Omega^+ \,.
\end{align}




From \eqref{et0}, 
we can conclude that \eqref{etN} holds. From this last consideration, the proof follows.
\hfill
\end{proof}

In the next section, we shall provide a sufficient condition for the FRS of 2D-systems; this result, together with Lemma \ref{FTSforILC}, will be exploited in Section \ref{section_ILC_solution} to provide an algorithm for the solution of Problem \ref{p1}. 

\section{A Sufficient Condition for the FRS of 2D-Systems} \label{sec:3}
The following result is a sufficient condition for the FRS of system \eqref{2Dsys}.

\begin{theorem}\label{t1}
Given the finite set~$\mathcal N_1\times \mathcal N_2$, the symmetric positive definite 2D matrix-valued sequence $R(h,k):=\diag\left(R_1(k),R_2(h)\right)$, and the symmetric positive definite 2D matrix-valued sequence $\Gamma(h,k):=$ $\diag\left(\Gamma_1(h,k),\Gamma_2(h,k)\right)$, with $\Gamma(0,0)<R(0)$, the discrete-time linear 2D-system~\eqref{2dsys_compact} is finite-region stable with respect to~$(\mathcal N_1\times \mathcal N_2,R(h,k),\Gamma(h,k))$,  if there exist a symmetric block-diagonal 2D matrix-valued sequence  $P(h,k)$ in the form \eqref{defP}, such that  
\begin{subequations} \label{conditions_main}
\begin{align}
  A^T(h,k)P^{+}(h,k)A(h,k)-P(h,k)&<0\,, \quad (h,k)\in \mathcal N_1 \times \mathcal N_2
  \label{classical_a}\\
  P(h,k) &>\Gamma (h,k)\,, \quad  (h,k)\in \mathcal N_1\times \mathcal N_2 \label{D2}\\
   {{P}_{1}}(0,k) &\le {{R}_{1}}(k)\,,\quad  k\in \mathcal N_2, \label{D3} \\
   {{P}_{2}}(h,0)&\le {{R}_{2}}(h)\quad h\in \mathcal N_1 \label{D4}\\
   P_1(N_1+1,k)& \geq 0\,, \quad k\in \mathcal N_2/\{N_2\} \label{D5} \\
   P_2(h,N_2+1)&\geq 0\,, \quad h\in\mathcal N_1/\{N_1\}\label{D6}\,,
   \end{align}
\end{subequations}
where 
$P^{+}(h,k)$ satisfies \eqref{defP+}.
\end{theorem}
\begin{proof}
Consider the symmetric block-diagonal 2D matrix-valued sequence  $P(h,k)$ in the form \eqref{defP}, and the corresponding quadratic Lyapunov function, evaluated along the trajectory $x(\cdot,\cdot)$,
\begin{align} \label{def_V_split}
 V(h,k)&:=x^T(h,k)P(h,k)x(h,k) \nonumber \\  
 & = x_1^T(h,k)P_1(h,k)x_1(h,k)
 + x_2^T(h,k)P_2(h,k)x_2(h,k)\nonumber \\
 &=: V_1(h,k)+V_2(h,k) \,.
\end{align}
\begin{figure}
\setlength{\unitlength}{1mm}
\begin{center}
\begin{picture}(70,60)
\put(-10,10){\vector(1,0){80}} 
\put(0,0){\vector(0,1){70}}
\put(-2,7){$0$} \put(8,7){$1$} \put(18,7){$\dots$}\put(28,7){$N_1$}\put(38,7){$\dots$}\put(48,7){$N_2$} \put(66,7){$k$}
\put(-2,17){$1$}\put(-2,27){$\vdots$}\put(-5,38){$N_1$}\put(-2,47){$\vdots$}\put(-5,57){$N_2$}\put(-4,65){$h$}
\put(0,39){\line(1,0){50}}\put(50,10){\line(0,1){29}}
\put(0,34){\line(1,-1){24}}\put(0,54){\line(1,-1){44}}\put(0,69){\line(1,-1){59}}
\put(10,19){$a$}\put(20,29){$b$}\put(30,35){$c$}

\end{picture}
\end{center}
\caption{The region $\mathcal N_1\times\mathcal N_2$, with $N_1< N_2$. Case a) $0<M\le N_1$; case b): $N_1 < M \le N_2$; case c): $N_2<M\le N_1+N_2$.} \label{Fig_Mvalue}
\end{figure}
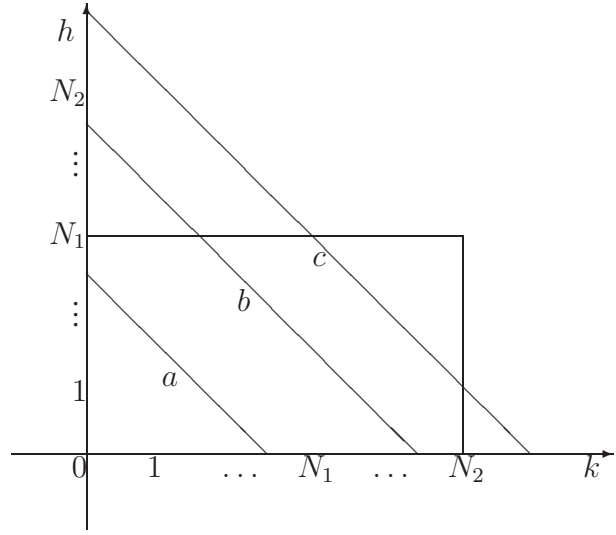

For all $(h,k)\in \mathcal N_1 \times \mathcal N_2$ 
the first difference of $V$, along the system trajectories, yields, in view of \eqref{def_V_compact} and \eqref{classical_a},
\begin{align}  \label{negative_definiteness}
 & V^+(h,k)-V(h,k) \nonumber \\
 &\hspace{-1cm} = x^T(h,k)\left(A^T(h,k)P^{+}(h,k)A(h,k) -P(h,k)\right)x(h,k) <0 \,.
\end{align}
Condition \eqref{negative_definiteness} implies, along the system trajectories,
\begin{align}  \label{negative_definiteness1}
  V^+(h,k)<V(h,k)\,,
\end{align}
for all $(h,k)\in\mathcal N_1\times \mathcal N_2$.

Now, observe that, for a given positive integer $M$, the points $(h,k)$ satisfying $h+k=M$ lie along the straight lines depicted in Figure \ref{Fig_Mvalue}.

According to Figure \ref{Fig_Mvalue} and without loss of generality, within the proof we shall assume that $N_1<N_2$.
In particular, there are three possible cases to consider: case a) when $0\le M < N_1 <N_2$; 
case b) when $N_1 \leq M < N_2$; 
case c) when $N_2\le M < N_1+N_2$. 

On the basis of the above discussion, in the following, we divide the proof into three parts, according to the value of $M$.

\noindent{}
{\bf Case a}, let $0\le M:=h+k < N_1 <N_2$. In this case, for any value of $M$, both $h$ and $k$ are strictly less than $N_1$ and $N_2$ respectively. Then, condition \eqref{def_V_compact}, \eqref{def_V_split} and \eqref{negative_definiteness1} return
\begin{equation}\label{iterative}
V_1(h+1,k)+V_2(h,k+1)  <    V_1(h,k)+
     V_2(h,k)\,,
\end{equation}
which, evaluated at the points $(M,0)$, $(M-1,1)$, $\cdots$, $(0,M)$, yields
\begin{subequations} \label{iterative_tot}
\begin{align}
V_{1}(M+1,0)+V_{2}(M,1)&< V_{1}(M,0)  +V_{2}(M,0)\label{iterative0}\\
V_{1}(M,1)+V_{2}(M-1,2)&<V_{1}(M-1,1)  
+V_{2}(M-1,1)\label{iterative1}\\
& \vdots \nonumber \\
V_{1}(1,M)+V_{2}(0,M+1)& <V_{1}(0,M) + V_{2}(0,M)\,.
 \label{iterativeM}
  \end{align}
\end{subequations}

Therefore, summing up the terms at the left and right hand side in \eqref{iterative_tot}, we obtain
\begin{equation}\label{eq_new}
\sum\limits_{h+k=M+1\atop{0\le M < N_1\atop{0\leq h\leq M+1}}}^{{}}{V(h,k)} < \sum\limits_{h+k=M\atop{0\le M < N_1\atop{0\leq h\leq M}}}^{{}}{V(h,k)}  
 +V_{1}(0,M+1)+V_{2}(M+1,0)\,.
\end{equation}

Letting $M\to M+1$ in \eqref{eq_new} and iterating, we have
\begin{align}
 \sum\limits_{h+k=M\atop{0<M\le N_1\atop{0\leq h\leq M}}}^{{}}{V(h,k)} &< \sum\limits_{h+k=M-1\atop{0<M\le N_1\atop{0\leq h\leq M-1}}}^{{}}{V(h,k)}   +V_{1}(0,M)+V_{2}(M,0) \nonumber \\ 
 & <\sum\limits_{h+k=M-2\atop{0<M\le N_1\atop{0\leq h\leq M-2}}}^{{}}{V(h,k)}+V_{1}(0,M-1) +V_{2}(M-1,0) \nonumber \\
 & \hspace{5cm}+V_{1}(0,M)+V_{2}(M,0) \nonumber \\
 &\vdots  \nonumber \\
 &< V(0,0)+V_{1}(0,1)+V_{2}^{T}(1,0)  \nonumber \\
 &\hspace{1cm} +\cdots +V_{1}(0,M-1) +V_{2}(M-1,0) \nonumber \\
 &\hspace{5cm}+V_{1}(0,M)+V_{2}(M,0) \nonumber \\
 &=\sum\limits_{k=0\atop {0<M\le N_1}}^{M}V_{1}(0,k)+\sum\limits_{h=0\atop {0<M\le N_1}}^{M}V_{2}(h,0)\label{k1_inter}\\
 &\leq \sum\limits_{k=0}^{N_2}V_{1}(0,k)+\sum\limits_{h=0}^{N_1}V_{2}(h,0). \label{k1}
\end{align}

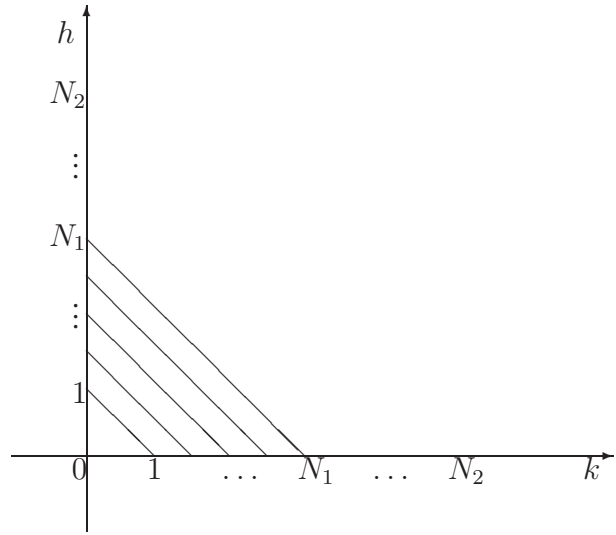
\begin{figure}
\setlength{\unitlength}{1mm}
\begin{center}
\begin{picture}(70,60)
\put(-10,10){\vector(1,0){80}} 
\put(0,0){\vector(0,1){70}}
\put(-2,7){$0$} \put(8,7){$1$} \put(18,7){$\dots$}\put(28,7){$N_1$}\put(38,7){$\dots$}\put(48,7){$N_2$} \put(66,7){$k$}
\put(-2,17){$1$}\put(-2,27){$\vdots$}\put(-5,38){$N_1$}\put(-2,47){$\vdots$}\put(-5,57){$N_2$}\put(-4,65){$h$}
\put(0,39){\line(1,-1){29}}\put(0,34){\line(1,-1){24}}\put(0,29){\line(1,-1){19}}\put(0,24){\line(1,-1){14}}\put(0,19){\line(1,-1){9}}

\end{picture}
\end{center}
\caption{The region $0\le h+k=M \le N_1$.} \label{fig_region_1}
\end{figure}

We have
\begin{eqnarray}\label{deriva}
1&\geq& \sum\limits_{k=0}^{N_2}{x_{1}^{T}(0,k){{R}_{1}}(k){{x}_{1}}(0,k)}+\sum\limits_{h=0}^{N_1}{x_{2}^{T}(h,0){{R}_{2}}(h){{x}_{2}}(h,0)}, \quad \text{in view of } \eqref{FTSdef} \nonumber \\
&\geq& \sum\limits_{k=0}^{N_2}{x_{1}^{T}(0,k){{P}_{1}}(0,k){{x}_{1}}(0,k)}+\sum\limits_{h=0}^{N_1}{x_{2}^{T}(h,0){{P}_{2}}(h,0){{x}_{2}}(h,0)},  \nonumber \\
&& \hspace{6cm} \text{in view of } \eqref{D3}- \eqref{D4} \nonumber \\
&>&  \sum\limits_{h+k=M\atop{0<M\le N_1\atop{0\leq h\leq M}}}^{{}}{V(h,k)}, \quad  \text{in view of } \eqref{k1} \nonumber \\
&\geq& x^T(h,k)P(h,k)x(h,k)\,, \quad \{(h,k): h+k=M \,, 0<M\le N_1  \} \nonumber \\
&>&  x^T(h,k)\Gamma(h,k)x(h,k) \quad \text{in view of } \eqref{D2}\nonumber \\
& &\hspace{2cm} \{(h,k): h+k=M \,, 0<M\le N_1  \}\,.
\end{eqnarray}
Therefore, we have proven the satisfaction of \eqref{FTS_cond_state} for all $(h, k)$, such that $0\le h+k\le N_1$, i.e. for all $h$ and $k$ which lay in the the triangular-shaped region depicted in Figure \ref{fig_region_1} (note that the condition \eqref{FTS_cond_state} is trivially satisfied for $h+k=0$, i.e. $h=0$, $k=0$, in view of the assumptions of the theorem).

\noindent{}
{\bf Case b}, let $ N_1 \leq M:=h+k < N_2$. In this case, for any value of $M$, $k$ is guaranteed to be strictly less than $N_2$, while the value of $h$ can be greater than $N_1$; hence we have to explicitly assume that $h \leq N_1$.


From condition \eqref{iterative} evaluated at the points $(N_1,M-N_1)$, $(N_1-1,M-N_1+1)$, $\cdots$, $(0,M)$, we obtain 
\begin{subequations}\label{itera3}
\begin{align} 
  V_1(N_1+1,M-N_1)+V_{2}({{N}_{1}},M-{{N}_{1}}+1) &
 <V_1(N_1,M-N_1)\nonumber \\
  & \hspace{0.5cm}+ V_{2}({{N}_{1}},M-{{N}_{1}}) \\
V_{1}({{N}_{1}},M-{{N}_{1}}+1)+ V_{2}({{N}_{1}}-1,M-{{N}_{1}}+2)
 & < V_{1}({{N}_{1}}-1,M-{{N}_{1}}+1) \nonumber\\
  & \hspace{0.5cm}+ V_{2}({{N}_{1}}-1,M-{{N}_{1}}+1) \\  
 & \hspace{2cm} \vdots \nonumber \\
  V_{1}(1,M)+V_{2}(0,M+1)&<V_{1}(0,M)  +V_{2}(0,M)\,. 
\end{align}
\end{subequations}

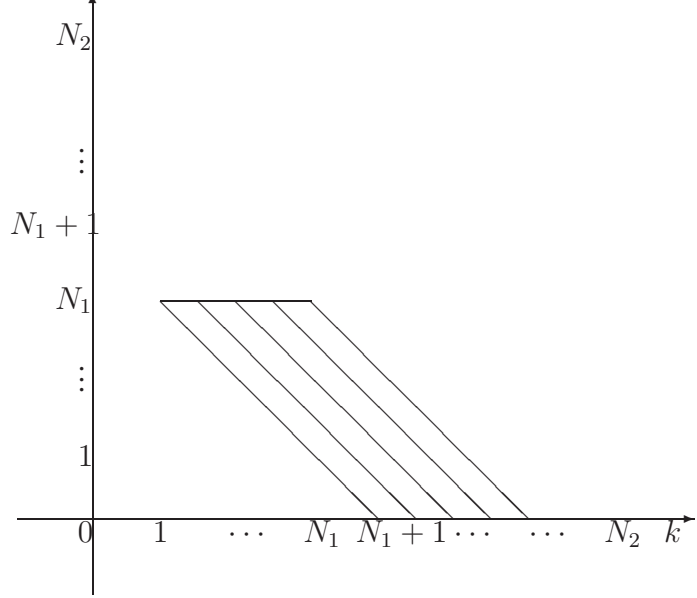
\begin{figure}
\setlength{\unitlength}{1mm}
\begin{center}
\begin{picture}(70,60)
\put(-10,10){\vector(1,0){90}} 
\put(0,0){\vector(0,1){80}}
\put(-2,7){$0$} \put(8,7){$1$} \put(18,7){$\cdots$}\put(28,7){$N_1$}\put(35,7){$N_1+1$}\put(48,7){$\cdots$} \put(58,7){$\cdots$}\put(68,7){$N_2$}\put(76,7){$k$}
\put(-2,17){$1$}\put(-2,27){$\vdots$}\put(-5,38){$N_1$}\put(-11,48){$N_1+1$}\put(-2,56){$\vdots$}\put(-5,73){$N_2$}\put(-4,80){$h$}
\put(9,39){\line(1,-1){29}}\put(14,39){\line(1,-1){29}} \put(19,39){\line(1,-1){29}}\put(24,39){\line(1,-1){29}}\put(29,39){\line(1,-1){29}}\put(9,39){\line(1,0){20}}
\end{picture}
\end{center}
\caption{The region $N_1\le h+k=M < N_2$.} \label{fig_region_2}
\end{figure}

Therefore, summing up the terms at the left and right hand side in \eqref{itera3},  
it yields
\begin{align}\label{eq_new_bis}
    \sum\limits_{h+k=M+1\atop{N_1\le M < N_2\atop{0\leq h\le N_1}}}^{{}}{V(h,k)}&< \sum\limits_{h+k=M\atop{N_1\leq M < N_2\atop{0\leq h\le N_1}}}^{{}}{V(h,k)} +V_{1}(0,M+1) \nonumber \\
&\hspace{3cm} -V_1(N_1+1,M-N_1))\,.
\end{align}

Letting $M\to M+1$ in \eqref{eq_new_bis}, and taking into account that the last term is nonpositive due to \eqref{D5}, 
we have: 
\begin{align}\label{k2_ante}
    \sum\limits_{h+k=M\atop{N_1< M \le N_2\atop{0\leq h\leq N_1}}}^{{}}{V(h,k)}&< \sum\limits_{h+k=M-1\atop{N_1< M \le N_2\atop{0\leq h\leq N_1}}}^{{}}{V(h,k)} +V_{1}(0,M)\,. 
\end{align}

Note that condition \eqref{k2_ante}, rewritten for $M=N_1+1$, yields
\begin{align} \label{k1_inter2}
    \sum\limits_{h+k=N_1+1\atop{0\leq h\leq N_1}} V(h,k) 
    &< \sum\limits_{h+k=N_1\atop{0\leq h\leq N_1}} V(h,k)+V_{1}(0,N_1+1)\,;
\end{align}
moreover, letting $M=N_1$ in condition \eqref{k1_inter}, we obtain
\begin{align}\label{k1_derived}
& \sum\limits_{h+k=N_1\atop{0\leq h\leq N_1}}V(h,k) 
< \sum\limits_{k=0}^{N_1}V_{1}(0,k) +\sum\limits_{h=0}^{N_1}V_{2}(h,0)\,. 
\end{align}

Therefore, iterating \eqref{k2_ante}, we obtain
\begin{align}\label{k2}
\sum\limits_{h+k=M\atop{N_1<M\leq N_2\atop{0\leq h\le N_1}}}^{{}}{V(h,k)}&< \sum\limits_{h+k=M-1\atop{N_1<M\leq N_2\atop{0\leq h\le N_1}}}^{{}}{V(h,k)} 
+V_{1}(0,M)\nonumber \\
 & <\sum\limits_{h+k=M-2\atop{N_1<M\leq N_2\atop{0\leq h\le N_1}}}^{{}}{V(h,k)}+V_{1}(0,M-1)+V_{1}(0,M) \nonumber \\
 & \vdots \nonumber \\
 & < \sum\limits_{h+k=N_1+1\atop{0\leq h\le N_1}} V(h,k)
 +\sum\limits_{k=N_1+2\atop{N_1<M\leq N_2}}^{M} V_1(0,k) \nonumber \\
 &< \sum\limits_{h+k=N_1\atop{0\leq h\leq N_1}} V(h,k)
 +\sum\limits_{k=N_1+1\atop{N_1<M\leq N_2}}^{M} V_1(0,k)\quad {\rm in \,  view \, of \, \eqref{k1_inter2}}\nonumber \\
 &<\sum\limits_{k=0}^{N_1}V_{1}(0,k) +\sum\limits_{h=0}^{N_1}V_{2}(h,0) \nonumber \\
 & \hspace{2cm} +\sum\limits_{k=N_1+1\atop{N_1<M\leq N_2}}^{M} V_1(0,k) \,, \quad {\rm in \, view \, of \, \eqref{k1_derived}}\nonumber \\
 & =\sum\limits_{k=0\atop{N_1<M\leq N_2}}^{M}V_{1}(0,k) +\sum\limits_{h=0}^{N_1}V_{2}(h,0)\nonumber \\
  &\leq \sum\limits_{k=0}^{N_2}V_{1}(0,k)+\sum\limits_{h=0}^{N_1}V_{2}(h,0)\,, 
\end{align}
which is the same as \eqref{k1}. Therefore, by following a derivation similar to \eqref{deriva}, we prove the satisfaction of \eqref{FTS_cond_state} for all $(h, k)$, such that
$N_1 < h+k \le N_2$, $h\leq N_1$, i.e. for all $h$ and $k$ which lay in the  parallelogram-shaped region depicted in Figure \ref{fig_region_2}.

\noindent{}
{\bf Case c}, let $N_2\leq M:=h+k < N_1+ N_2$. In this case, according to Figure \ref{Fig_Mvalue}, for a given $M$, a point on the $M$-line can either stay inside the region $\mathcal N_1\times \mathcal N_2$, i.e. $h\leq N_1$ and $k\leq N_2$, 
or lies outside the above region, i.e. $h>N_1$ or $k>N_2$. Hence we have to explicitly assume that $h \leq N_1$ and $k\le N_2$.

Again, condition
\eqref{iterative} evaluated at the points $(N_1,M-N_1)$, $(N_1-1,M-N_1+1)$, $\cdots$, $(M-N_2,N_2)$, returns
\begin{subequations}\label{first_condition_last}
\begin{align} 
  V_1(N_1+1,M-N_1)+V_{2}({{N}_{1}},M-{{N}_{1}}+1)&
 <V_1(N_1,M-N_1)\nonumber \\
 & \hspace{0.5cm}+ V_{2}({{N}_{1}},M-{{N}_{1}}) \\
V_{1}({{N}_{1}},M-{{N}_{1}}+1) + V_{2}({{N}_{1}}-1,M-{{N}_{1}}+2)
 & < V_{1}({{N}_{1}}-1,M-{{N}_{1}}+1) \nonumber \\
 &\hspace{0.5cm}+ V_{2}({{N}_{1}}-1,M-{{N}_{1}}+1) \\ 
 & \vdots \nonumber \\
  V_{1}(M-N_2+1,N_2) + V_{2}(M-N_2,N_2+1)
 & < V_{1}(M-N_2,N_2) \nonumber \\
 &\hspace{0.5cm}+ V_{2}(M-N_2,N_2)\,.
\end{align}
\end{subequations}

Therefore, summing up the terms at the left and right hand side in \eqref{first_condition_last}, and taking into account conditions \eqref{D5}-\eqref{D6}, we have 
\begin{align} \label{free}
    \sum_{h+k=M+1\atop{{N_2\leq M < N_1+N_2}\atop{M-N_2+1\le h \le N_1}}} V(h,k) &< \sum_{h+k=M\atop{{N_2\leq M < N_1+N_2}\atop{M-N_2\le h \le N_1}}} V(h,k) -V_1(N_1+1,M-N_1) \nonumber \\
  &\hspace{3cm} - V_2(M-N_2,N_2+1)\nonumber \\
    &\leq \sum_{h+k=M\atop{{N_2\leq M < N_1+N_2}\atop{M-N_2\le h \le N_1}}} V(h,k)\,.
\end{align}

Letting $M\to M+1$ in \eqref{free}, we have:
\begin{equation}\label{free2}
     \sum_{h+k=M\atop{{N_2 < M \leq N_1+N_2}\atop{M-N_2\le h \le N_1}}} V(h,k) < \sum_{h+k=M-1\atop{{N_2< M \leq N_1+N_2}\atop{M-N_2-1\le h \le N_1}}} V(h,k)\,.
\end{equation}
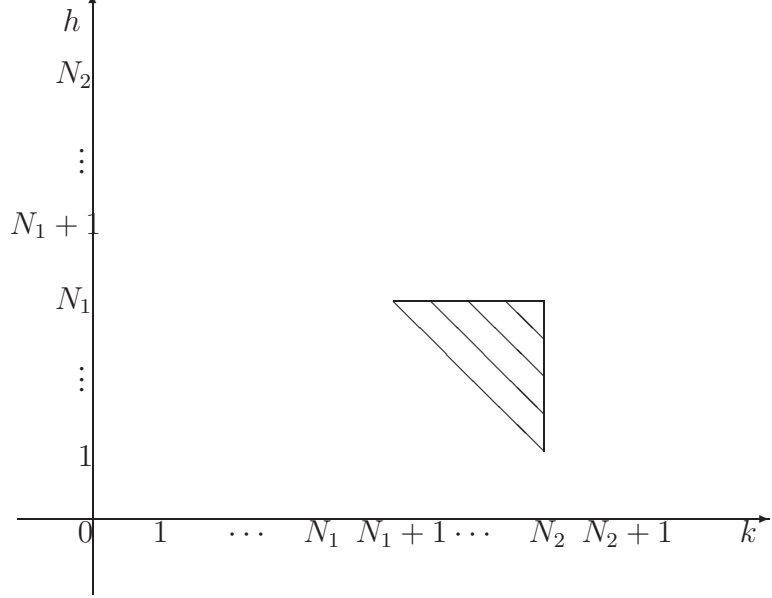
\begin{figure}
\setlength{\unitlength}{1mm}
\begin{center}
\begin{picture}(70,60)
\put(-10,10){\vector(1,0){100}} 
\put(0,0){\vector(0,1){80}}
\put(-2,7){$0$} \put(8,7){$1$} \put(18,7){$\cdots$}\put(28,7){$N_1$}\put(35,7){$N_1+1$}\put(48,7){$\cdots$} \put(58,7){$N_2$}\put(65,7){$N_2+1$}\put(86,7){$k$}
\put(-2,17){$1$}\put(-2,27){$\vdots$}\put(-5,38){$N_1$}\put(-11,48){$N_1+1$}\put(-2,56){$\vdots$}\put(-5,68){$N_2$}\put(-4,75){$h$}
\put(40,39){\line(1,-1){20}}
\put(45,39){\line(1,-1){15}}
\put(50,39){\line(1,-1){10}}
\put(55,39){\line(1,-1){5}}
\put(40,39){\line(1,0){20}}
\put(60,39){\line(0,-1){20}}

\end{picture}
\end{center}
\caption{The region $N_2<M\le N_1+N_2$.} \label{fig_region_3}
\end{figure}

Condition \eqref{free2}, rewritten for $M=N_2+1$, yields 
\begin{equation} \label{free3}
    \sum_{h+k=N_2+1\atop{{1\leq h\leq N_1}}} V(h,k) < \sum_{h+k=N_2\atop{{0\leq h\leq N_1}}} V(h,k)\,.
    \end{equation}
    
Now, rewriting condition \eqref{k2}  for $M=N_2$, we have
\begin{equation}\label{free4}
    \sum_{h+k=N_2\atop{{0\leq h\leq N_1}}}{V(h,k)} <\sum\limits_{k=0}^{N_2}V_{1}(0,k)+\sum\limits_{h=0}^{N_1}V_{2}(h,0)\,.
\end{equation}

Therefore, iterating \eqref{free2}, we obtain
\begin{align}
    \sum_{h+k=M\atop{{N_2< M\leq N_1+N_2}\atop{M-N_2\le h \le N_1}}} V(h,k) &< \sum_{h+k=M-1\atop{{N_2< M \leq N_1+N_2}\atop{M-N_2-1\le h \le N_1}}} V(h,k)\nonumber \\
    &<\sum_{h+k=M-2\atop{{N_2< M \leq N_1+N_2}\atop{M-N_2-2\le h \le N_1}}} V(h,k)\nonumber \\
    &\vdots \nonumber \\
    &<\sum_{h+k=N_2+1\atop{{1\leq h \leq N_1}}} V(h,k)\nonumber \\
    &<\sum_{h+k=N_2\atop{{0\leq h \leq N_1}}} V(h,k) \,, \quad {\rm in \, view\, of \, \eqref{free3}}\nonumber \\
    &<\sum\limits_{k=0}^{N_2}V_{1}(0,k)+\sum\limits_{h=0}^{N_1}V_{2}(h,0) \quad {\rm in \,view\, of\, \eqref{free4}}\,,
\end{align}
which, again, is the same as \eqref{k1}. Therefore, by following a derivation similar to \eqref{deriva}, we prove the satisfaction of \eqref{FTS_cond_state} for all $(h, k)$, such that
$N_2<h+k\le N_1+N_2$, $h\leq N_1$, $k\leq N_2$, i.e. for all $h$ and $k$ which lay in the  triangular-shaped region depicted in Figure \ref{fig_region_3}. This completes the proof.
\hfill
\end{proof}

\begin{remark}
As it clearly appears from direct comparison, the FRS proof for linear 2D-systems requires a much more complex machinery than the corresponding FTS proof for 1D-systems (see \cite{FTS_book}). 
Also, due to the finiteness of the region $\mathcal N_1\times \mathcal N_2$, the proof of Theorem \ref{t1} is more challenging than the proof of classical Lyapunov stability for 2D-systems (see \cite{Kakzorek}), since, in the last case, the radial unboundedness of the domain simplifies the mathematical derivation. 
\end{remark}
\begin{remark} \label{rem_LMIs}
The application of Theorem \ref{t1} requires the solution of a feasibility problem constrained by the DLMI \eqref{classical_a}, and the set of LMIs obtained from \eqref{D2}--\eqref{D6}. In turn, the DLMI \eqref{classical_a} can be implemented through a set of LMIs, one for each value of $h$ and $k$. Hence, contrarily to the continuous-time case (see \cite{FTS_book}, Ch. 2), DLMIs naturally lead to a set of LMIs, 
and, therefore, the numerical implementation of the conditions in Theorem \ref{t1} does not add conservativeness to the FRS  test.
\hfill$\blacksquare$
\end{remark}
\subsection{Comparison with the Previous Literature}\label{sec:ex_comparison}
There are many improvements introduced by Theorem \ref{t1}, with respect to the previous literature concerning the FRS of 2D-systems. In previous papers \cite{Zhang_2016,Hua_2018,Hua_2018_b,Gao}, linear time-invariant systems have been dealt with, and an approach based on time-invariant quadratic Lyapunov functions was exploited. Conversely, the approach of Theorem \ref{t1}, which exploits time-varying Lyapunov functions, allows: i) to deal with the more general class of linear time-varying systems; ii) to consider, in Definition \ref{def_FSS}, ellipsoidal-shaped initial and trajectories domains varying over the region of interest, which is a definitely more general context,  closer to real engineering problems, than the rough bound $x^T\Gamma x \leq c$, with $c$ scalar, considered in previous papers (note, for instance, that the works concerning the FRS of 2D-systems published in the previous literature, could not be used to develop ILC techniques); 
iii) to obtain less conservative conditions for FRS, since the class of time-invariant Lyapunov functions is a small subset of the class of time-varying functions.  

To better illustrate the point ii), let us consider the following example.

\begin{example}\label{example-comparison}
Let us consider system \eqref{2dsys_compact}, with
\begin{subequations}
\begin{align}
N_1&=10\,,\quad N_2=10\\
R_1(k)&=R_2(h)=I\\
\Gamma_1(h,k)& = \Gamma_2(h,k) = 0.8\times 1.22^{h+k}\,I\,,
\end{align}
and
\begin{equation}
    A(h,k):=\begin{pmatrix}
  0.2 & 0.5-0.01h \\
    0.2+0.02k & 0.2
\end{pmatrix}\,.
\end{equation}
\end{subequations}

The optimization problem constrained by conditions \eqref{conditions_main}, has been implemented in Matlab$^{\Registered}$, exploiting YALMIP \cite{Lofberg2004} and the MOSEK optimization toolbox \cite{mosek}. For this case, the optimization algorithm yields a solution (the computation only requires a few seconds on a standard desktop computer), i.e. the problem is feasible, which means the system is FRS with respect to $(\mathcal N_1 \times \mathcal N_2,R(h,k),\Gamma(h,k))$, for all $(h,k) \in \mathcal N_1\times \mathcal N_2$.  

Figure \ref{fig:ex1_funnel} displays both the initial condition bound expressed by \eqref{FTSdef} and the trajectory bound given by \eqref{FTS_cond_state}. Note that, for the well-posedness of the problem, the trajectory bound must contain the initial condition bound; however, leveraging to the space-varying formulation proposed in this paper, the bound can be exponentially shrunk (or enlarged) when $h$ and $k$ increase. Other approaches, previously appeared in literature \cite{Zhang_2016,Hua_2018,Hua_2018_b,Gao}, could only deal with FRS when $\Gamma(\cdot,\cdot)$ is constant for all $(h,k)$ over the interval $\{0,\dots,10\}\times \{0,\dots,10\}$; this would entail a fixed bound like the one reported in Figure \ref{fig:ex1_funnel}.

\begin{figure}
  \centering
  \includegraphics[width=12cm]{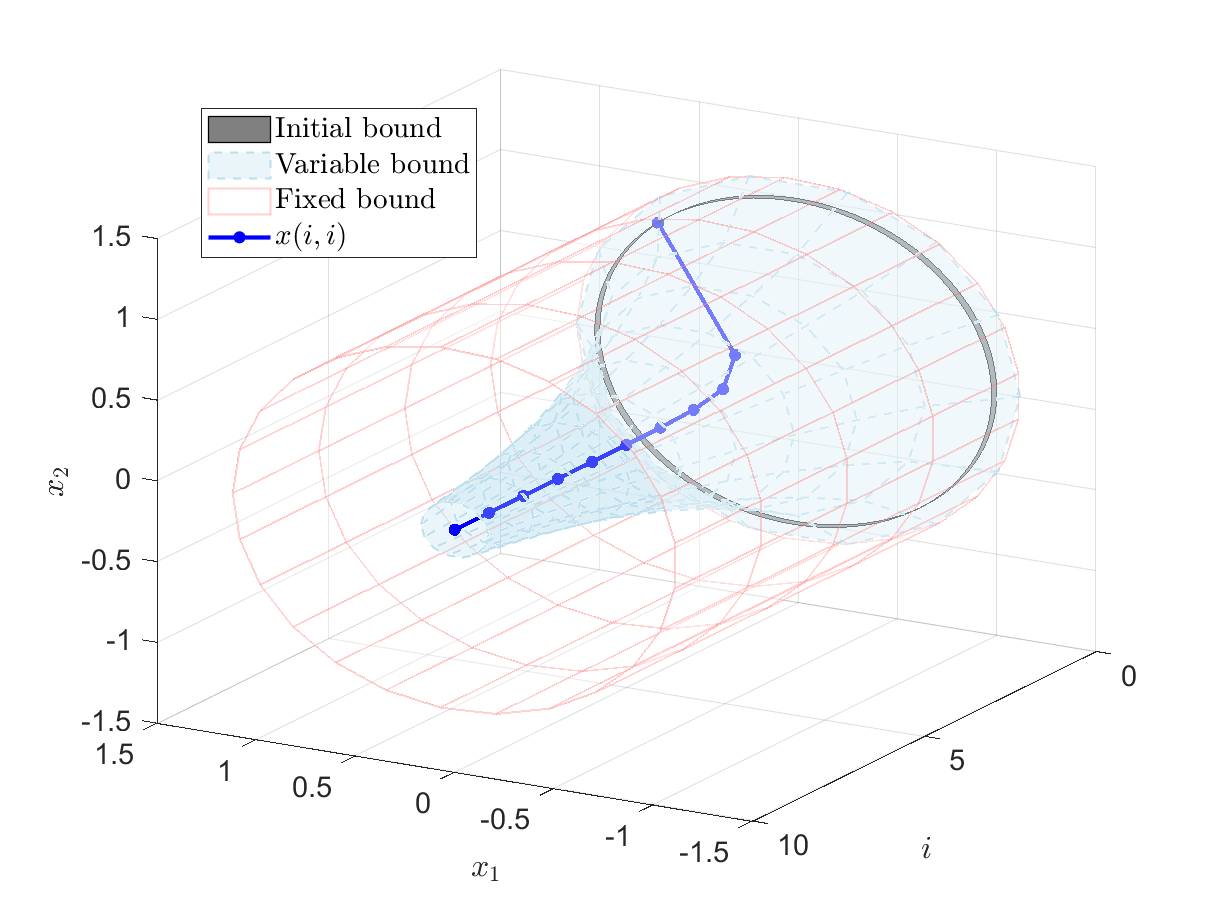}
  \caption{Example 1: the initial condition bound, defined by $R_1(k),R_2(h)$, must be surrounded by the trajectory bound, defined by  $\Gamma_1(\cdot,0), \Gamma_2(0,\cdot)$. Using a variable weighting matrix $\Gamma(h,k)$ it is possible to shrink or enlarge the bound over $\mathcal N_1\times \mathcal N_2$. Using a constant $\Gamma$ implies a fixed bound for all $(h,k)$. To graphically render the above concepts, only the diagonal point ($h=k$) have been considerd in the figure.}\label{fig:ex1_funnel}
\end{figure}

\begin{figure}
  \centering
  \includegraphics[width=10cm,height=8cm]{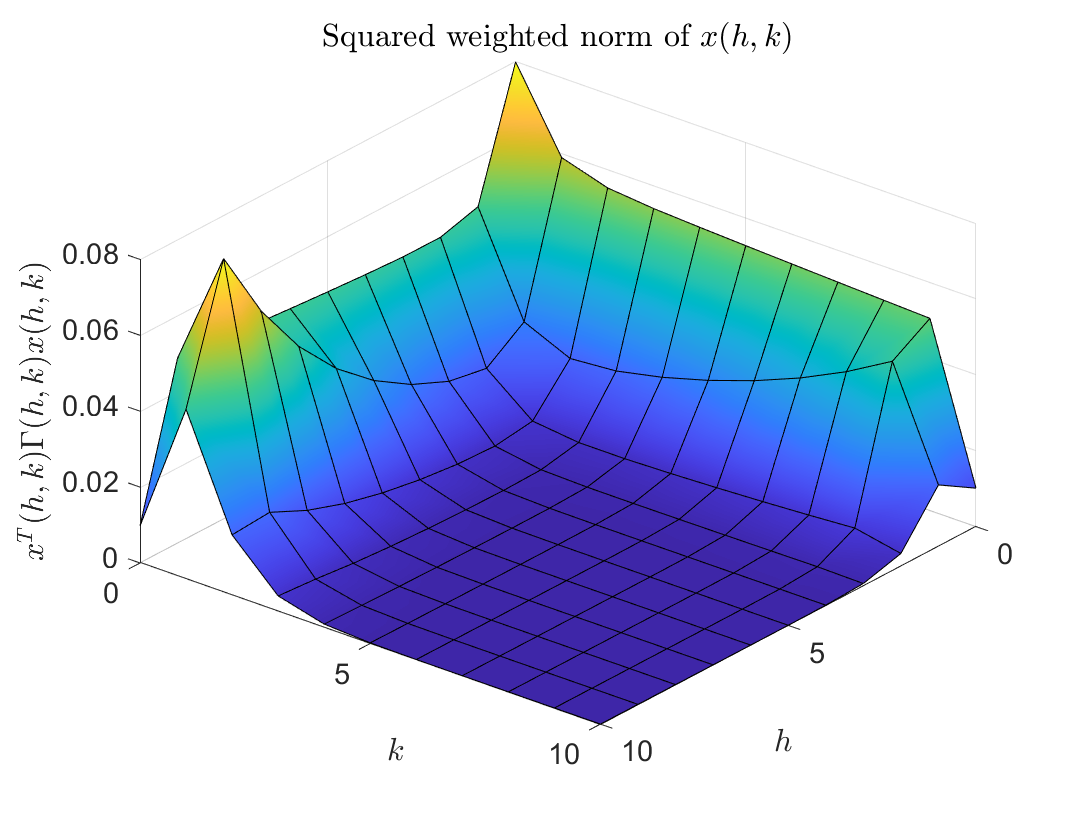}
  \caption{Example 1 -- the weighted norm of the state over the region $\mathcal N_1 \times \mathcal N_2$, with initial condition $x1(1,k) = x2(h,1) = 0.2236$.}\label{fig:ex2_funnel}
\end{figure}

\end{example}

 In the next section, we shall illustrate an interesting application of the FRS theory in the context of ILC.

\subsection{ILC Design via LMIs}\label{section_ILC_solution}
The next result exploits Lemma \ref{FTSforILC} and Theorem \ref{t1} in order to solve Problem \ref{p1} in a way suitable of numerical implementation.

\begin{theorem} \label{th_ILC}
Problem \ref{p1} is solvable if
there exist a 2D symmetric positive definite matrix-valued sequence $Q(t,k)=\diag(Q_1(t,k),Q_2(t,k))$, and a  2D matrix-valued sequence $L(t,k)$, such that (the argument $(t,k)$ 
is omitted for brevity)
\begin{subequations}
\begin{align}
\label{cond_sf_a_ILC}    \begin{pmatrix}
    -Q & Q\left(A_{\text{ILC}}^{\text{OL}}\right)^T + \tilde L^T\left(B_{\text{ILC}}^{\text{OL}}\right)^T\\
  A_{\text{ILC}}^{\text{OL}}Q+B_{\text{ILC}}^{\text{OL}}
  \tilde L          & -Q^{+}
  \end{pmatrix} &<0, \quad (t,k)\in \Omega^+\times \mathcal N\\
\label{cond_sf_e_ILC} 
&\hspace{-1cm} Q_2< \tilde{\Gamma}^{-1}\,,\quad (t,k)\in\Omega^+\times \mathcal N\\
\label{cond_sf_g_ILC}   Q_2(t,0)&\geq \tilde{R}^{-1}\,, \quad  t\in  \Omega^+ \\
\label{cond_sf_g_ILC_Q1}
Q_1(T+1,k)&>0 \,, \quad k\in\mathcal N/\{N\}\\
\label{cond_sf_g_ILC_Q2}
Q_2(t,N+1) &>0 \,, \quad t\in\Omega^+/\{T\}\,,
\end{align}
\end{subequations}
where
\begin{subequations}
\begin{align}
    A_{\text{ILC}}^{\text{OL}}(t)&:=\begin{pmatrix}
    A(t-1) & 0 \\
    -C(t)A(t-1) & I
    \end{pmatrix}\,, \quad B_{\text{ILC}}^{\text{OL}}(t):=\begin{pmatrix}
    B(t-1) \label{clA_ILC} \\
    -C(t)B(t-1)
    \end{pmatrix} \\
    \tilde L(t,k)&:= \begin{pmatrix} 0 & L(t,k) \end{pmatrix} \label{clL_ILC} \,, 
\end{align}
\end{subequations}

In this case, a finite-region stabilizing control law has the form \eqref{update}, with $K(t,k)=L(t+1,k)Q_2^{-1}(t+1,k)$.
\end{theorem}
\begin{proof}
According to Lemma \ref{FTSforILC}, Problem \ref{p1} is solvable if
there exist scalars $\epsilon_1 >\epsilon_2>0$, such that system \eqref{Roesser} is finite-region stable with respect to $\left(\Omega^+\times \mathcal N\right.$, $\left. \diag(\epsilon_1 I,\tilde{R}),\diag(\epsilon_2 I, \tilde{\Gamma}(t,k)\right)$.

Therefore, in view of Theorem \ref{t1}, Problem \ref{p1} is solvable if
there exist scalars $\epsilon_1>\epsilon_2 >0$, and symmetric positive definite matrix-valued sequences  $P_1(t,k)$, $P_2(t,k)$, of compatible dimensions, such that
\begin{subequations}
\begin{align}
A^T_{\text{ILC}}(t,k)P^{+}(t,k)A_{\text{ILC}}(t,k)-P(t,k)&<0\,, \quad (t,k)\in \Omega^+\times \mathcal N \label{classical_a_ILC}\\
P_1(t,k) &> \epsilon_2 I, \quad (t,k)\in \Omega^+ \times \mathcal N \label{D2_ILC_1} \\
P_2(t,k) &>\tilde{\Gamma} (t,k),\quad (t,k)\in \Omega^+ \times \mathcal N  \label{D2_ILC_2}\\
{{P}_{1}}(1,k) &\le \epsilon_1 I , \quad k\in \mathcal N, \label{D2_ILC_3} \\
  {{P}_{2}}(t,0)&\le \tilde{{R}}, \quad t\in \Omega^+,\label{D2_ILC_4}\\
\label{D2_ILC_last1} P_1(T+1,k)& \geq 0\,, \quad k\in \mathcal N/\{N\}\\
\label{D2_ILC_last} P_2(t,N+1)& \geq 0\,, \quad t\in\Omega^+/\{T\}\,,
   \end{align}
\end{subequations}
where $P(t,k)=\diag\bigl(P_1(t,k),P_2(t,k)\bigr)$.


 By pre- and post-multiplying condition \eqref{classical_a_ILC} by $P^{-1}(t,k)=:Q(t,k)$, we obtain the equivalent DLMI 
\begin{equation}\label{classical_a_sf_1_ILC}
     Q(t,k)A_{{\text{ILC}}}^T(t,k)Q^{+^{-1}}(t,k)A_{{\text{ILC}}}(t,k)Q(t,k)-Q(t,k)<0 \,.
\end{equation}

Now, exploiting \eqref{clA_ILC}, we can write
\begin{equation}
    A_{ILC}(t,k)=A_{ILC}^{OL}(t)+B_{ILC}^{OL}(t) \begin{pmatrix}
       0 & K(t-1,k)
    \end{pmatrix}\,;
\end{equation}
therefore, by using Shur complements arguments and \eqref{clL_ILC}, inequality \eqref{classical_a_sf_1_ILC} can be rewritten (we omit the argument $(t,k)$ for the sake of simplicity)
    \begin{align} \label{cond_a_ILC}
 \begin{pmatrix}
    -Q & QA_{\text{ILC}}^T \\
  A_{\text{ILC}}Q   & -Q^{+}
  \end{pmatrix}
 &= \begin{pmatrix}
    -Q & Q\left(A_{\text{ILC}}^{\text{OL}}\right)^T + \tilde L^T\left(B_{\text{ILC}}^{\text{OL}}\right)^T\\
  A_{\text{ILC}}^{\text{OL}}Q+B_{\text{ILC}}^{\text{OL}}
  \tilde L          & -Q^{+}
  \end{pmatrix}\nonumber \\
     &<0\,,
    \end{align}\,.
 
 From this last consideration, we have that condition \eqref{classical_a_ILC}, i.e. \eqref{cond_a_ILC}, is equivalent to \eqref{cond_sf_a_ILC}. By using the properties of symmetric positive definite matrices and the fact that $Q(t,k)= \diag\bigl(Q_1(t,k),Q_2(t,k)\bigr):=$ $\diag\bigl(P_1^{-1}(t,k),P_2^{-1}(t,k)\bigr)$, it is readily seen that condition \eqref{D2_ILC_1} is satisfied by taking any scalar $\epsilon_2$ such that
\begin{equation}
    \epsilon_2 < \min_{(t,k)\in\Omega^+\times \mathcal N} {\frac{1}{\lambda_{\max} \left(Q_1(t,k)\right)}}\,;
\end{equation}
and, in the same way, condition \eqref{D2_ILC_3} is satisfied by letting
\begin{equation}
    \epsilon_1\geq \max_{k\in \mathcal N}{\frac{1}{\lambda_{\min} \left(Q_1(1,k)\right)}}\,,
\end{equation}
where $\lambda_{\min}(\cdot)$ and $\lambda_{\max}(\cdot)$ denote the minimum and maximum eigenvalues of the argument respectively.

Finally, it is simple to recognize that conditions \eqref{cond_sf_e_ILC} and \eqref{cond_sf_g_ILC} are equivalent to \eqref{D2_ILC_2} and \eqref{D2_ILC_4}, and that conditions \eqref{cond_sf_g_ILC_Q1}, \eqref{cond_sf_g_ILC_Q2} in turn imply \eqref{D2_ILC_last1}, \eqref{D2_ILC_last}.~\hfill
\end{proof}

In the next section, a numerical example will illustrate the application of Theorem \ref{th_ILC}.

\subsection{An Illustrative Example}\label{sec:4_3}
In this section we illustrate the application of Theorem 2 to design an iterative learning controller for the example system
\begin{subequations}\label{eq:ex_sec_33}
\begin{align}
x(t+1) &= 
    \begin{pmatrix}
        0.4075 + 0.2037\cos(\frac{2\pi}{10}t)  &   0.2528\\
        0.2528  &   -0.445
    \end{pmatrix} \, x(t) +
    \begin{pmatrix}
        -0.7236\\
        -0.5933
    \end{pmatrix} \, u(t) \\
    y(t) &= \begin{pmatrix}
        0.4013  &   0.9421
    \end{pmatrix} \, x(t)
\end{align}
\end{subequations}
The goal is to design a controller that guarantees tracking of the reference signal $y_r(t) = 2\sin(0.3 \,t)$ over the interval $\Omega=\{0,\dots,20\}$. 
The number of iterations is fixed to $N=6$ and the weighting matrix has been set to 
\begin{equation*}
\tilde{\Gamma}(t,k) = \bar{\Theta} \left(1 + \rho^N - \rho^k \right)\,,
\end{equation*}
with $\bar{\Theta}=I$ and $\rho = 0.8517$. Note that, with this choice of the $\tilde{\Gamma}(t,k)$ function, the weight is increased up to $\bar{\Theta}$ at each ILC iteration $k$ (thus we expect the error to decrease, accordingly), whereas at each iteration the same weight is assigned to the tracking error at each time point.
\begin{figure}[ht]
  \centering
  \includegraphics[width=8cm]{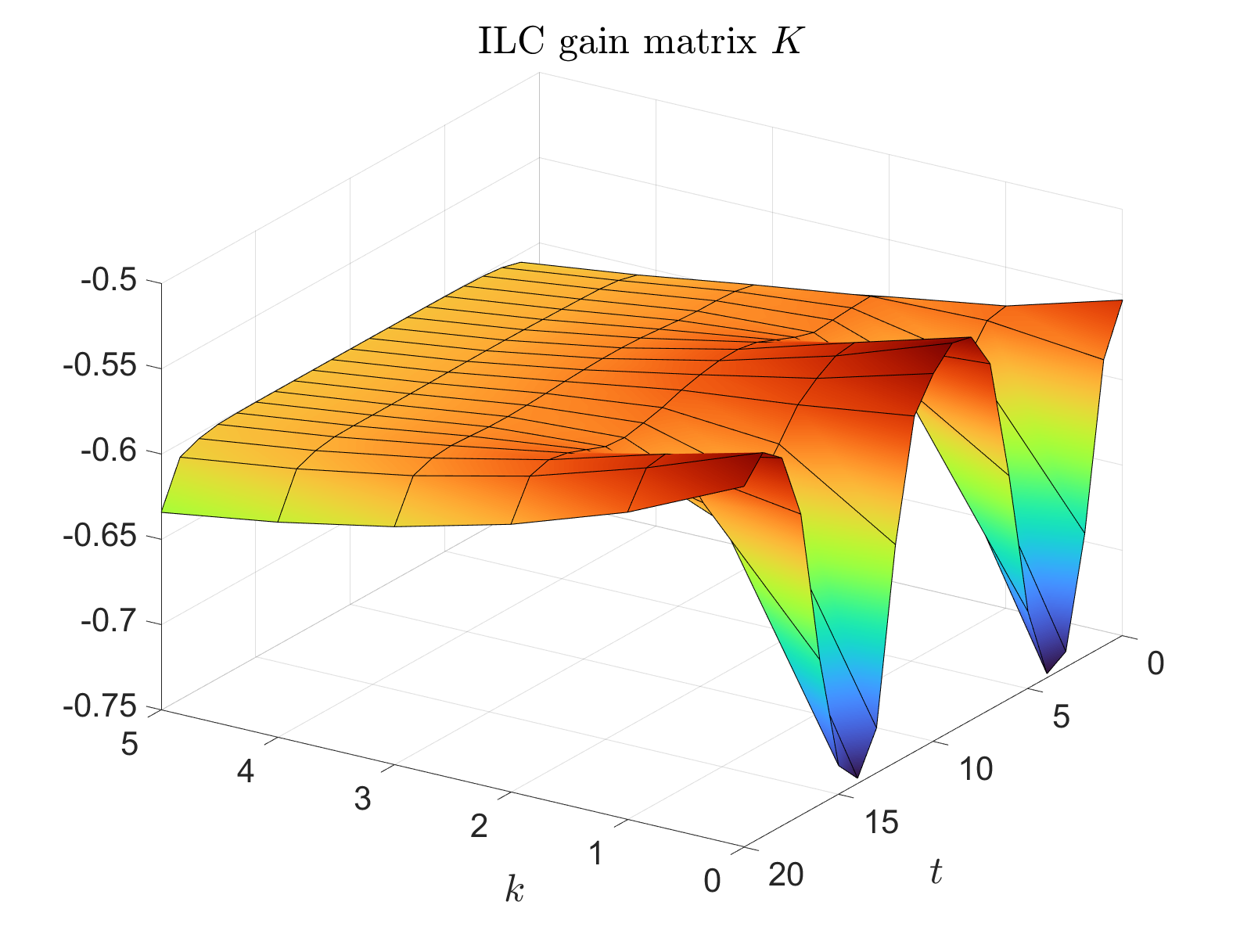}\\
  \caption{Controller $K(t,k)$ designed for system~\eqref{eq:ex_sec_33} leveraging the conditions of Theorem~\eqref{th_ILC}.}\label{fig:K_ILC}
\end{figure}

\begin{figure}[ht]
  \centering
  \includegraphics[width=8cm]{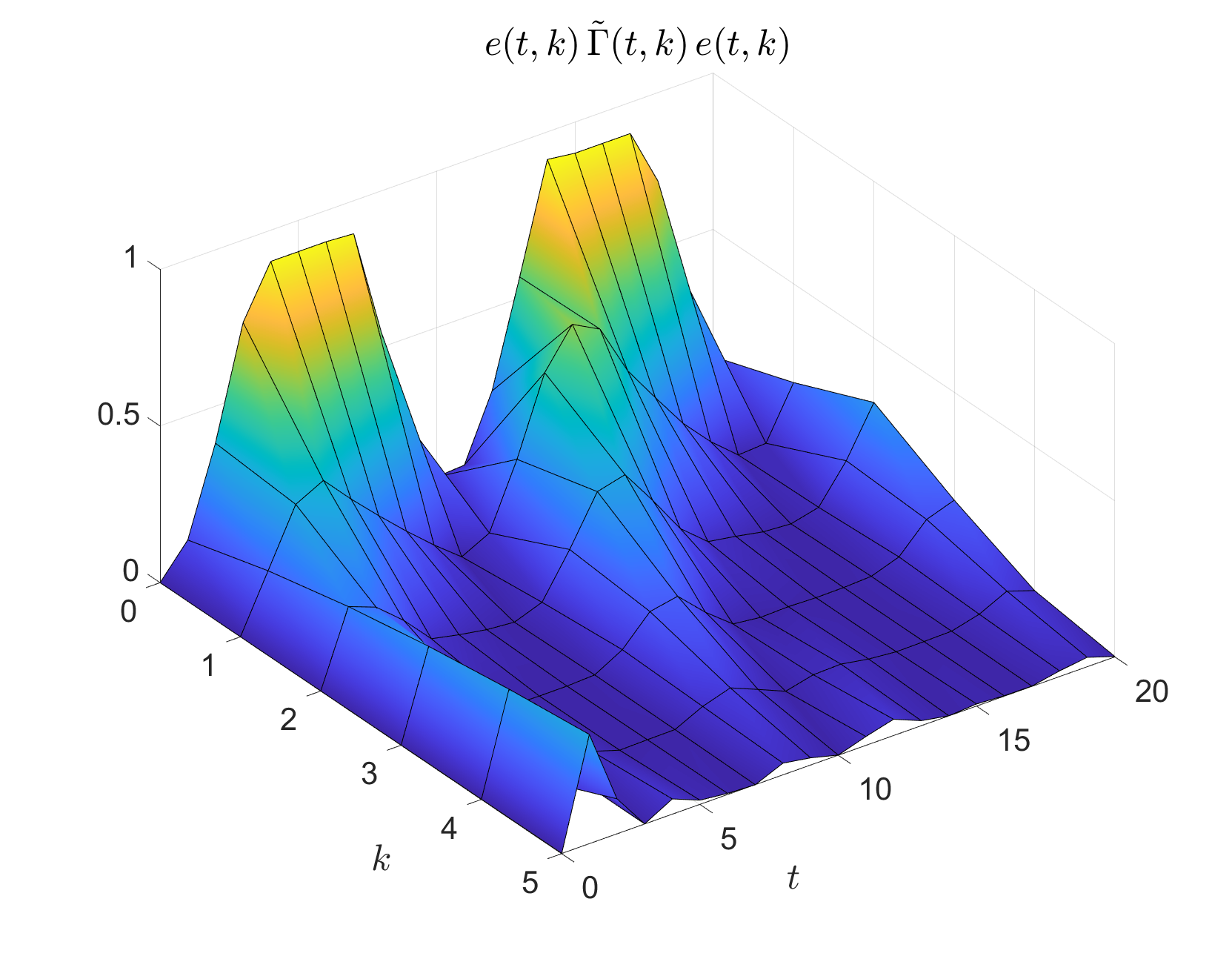}\\
  \caption{Weighted norm of the closed-loop system formed by \eqref{eq:ex_sec_33} with the iterative learning controller $K(t,k)$ shown in Fig.~\ref{fig:K_ILC}.}\label{fig:ILC_weighted_norm}
\end{figure}
\begin{figure}[ht]
  \centering
  \includegraphics[width=8cm]{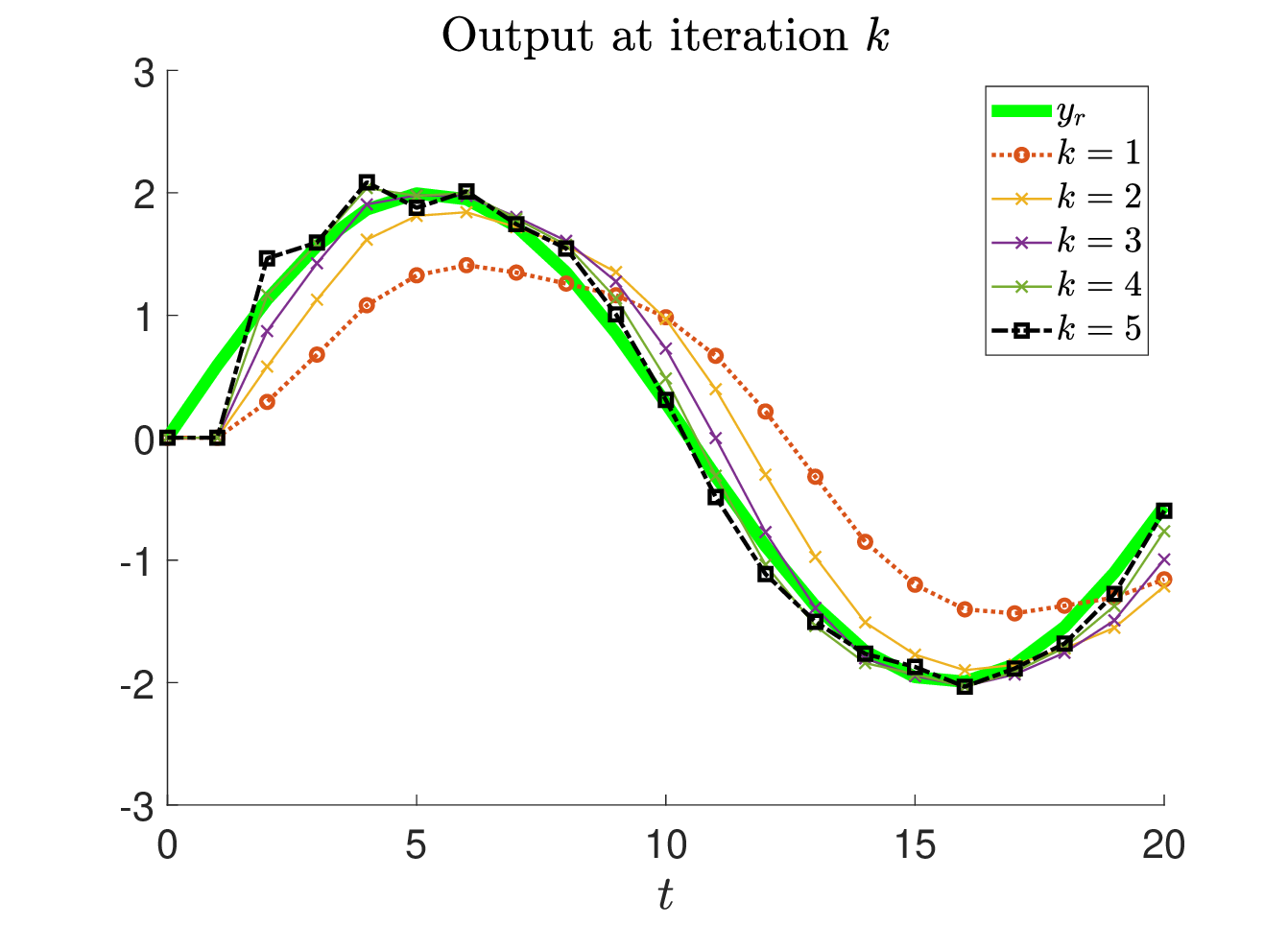}\\
  \caption{Output of the closed-loop system formed by \eqref{eq:ex_sec_33} with the iterative learning controller $K(t,k)$ shown in Fig.~\ref{fig:K_ILC}.}\label{fig:ILC_y}
\end{figure}
The conditions of Theorem~\eqref{th_ILC} applied to system~\eqref{eq:ex_sec_33}, yield the controller $K(t,k)$ depicted in Fig.~\ref{fig:K_ILC}, whereas Fig.~\ref{fig:ILC_weighted_norm} reports the weighted norm of the closed-loop system obtained by applying the designed iterative learning controller to system~\eqref{eq:ex_sec_33}.

The evolution of the output of the closed-loop system is reported in Fig.~\ref{fig:ILC_y}: it can be noticed that the trajectory approaches the desired reference as $k \rightarrow N$, as expected by an ILC controller.
As in section \ref{sec:ex_comparison}, the DLMI problem has been cast and solved in Matlab, exploiting YALMIP and MOSEK toolboxes, and the solution has required just a few seconds on a standard desktop computer.

\FloatBarrier

\section{Finite-Region Stabilization}\label{sec:2_b}
In this section, we extend the results of Section \ref{sec:3} to find a general state feedback control law for system \eqref{2dsys_compact}.
To this end, let us consider system \eqref{2dsys_compact} with a control input
\begin{equation} \label{2dsys_compact_u}
    x^{+}(h,k) = A(h,k) x(h,k) + B(h,k) u(h,k)\,,
\end{equation}
where
\begin{equation}
    B(h,k)=\begin{pmatrix}
    B_{1}(h,k)\\
    B_{2} (h,k)
    \end{pmatrix}\,,
\end{equation}
and $B(h,k)$ is partitioned according to  \eqref{2Dsys}.

The finite-region stabilization problem can be formalized as follows.
\begin{problem}\label{prob_sf}
Given the finite set $\mathcal N_1\times \mathcal N_2$, the symmetric positive definite 2D matrix-valued sequence $R(h,k):=\diag\left(R_1(k),R_2(h)\right)$, and the symmetric positive definite 2D matrix-valued sequence $\Gamma(h,k):=\diag\left(\Gamma_1(h,k),\Gamma_2(h,k)\right)$, with $\Gamma(0,0)<R(0,0)$, $\Gamma_1(0,k)<R_1(k)$, $k\in\mathcal N_2$, $\Gamma_2(h,0)<R_2(h)$, $h\in\mathcal N_1$, find a state feedback control law in the form
\begin{equation}\label{state_feedback}
    u(h,k)=K(h,k) x(h,k)\,,
\end{equation}
where
\begin{equation}
    K(h,k)=\begin{pmatrix}
    K_1(h,k) &
    K_2(h,k)
    \end{pmatrix}
\end{equation}
is partitioned according to  \eqref{2Dsys}, such that the closed loop system, given by the connection of system \eqref{2dsys_compact_u} and the control law \eqref{state_feedback}, is finite-region stable with respect to $(\mathcal N_1 \times \mathcal N_2,R(h,k),\Gamma(h,k))$.
\end{problem}

The following result is a sufficient condition for the solvability of Problem \ref{prob_sf}.
\begin{theorem}\label{t3}
Problem \ref{prob_sf} is solvable if there exist a symmetric positive definite 2D matrix-valued sequence $Q(h,k)=\diag(Q_1(h,k),Q_2(h,k))$, and a 2D matrix-valued sequence $L(h,k)$, such that (the argument $(h,k)$ is omitted for brevity)
\begin{subequations}
\begin{align}
\label{cond_sf_a}    \begin{pmatrix}
    -Q & Q A^T  +  L^TB^T \\
  A Q +B L          & -Q^{+}
  \end{pmatrix} &<0\,, \quad (h,k)\in \mathcal N_1\times \mathcal N_2\\
\label{cond_sf_d}  \hspace{-4cm}   Q &< \Gamma^{-1} \,,\quad (h,k)\in \mathcal N_1\times \mathcal N_2,\\
\label{cond_sf_e}  \hspace{-4cm} Q_1(0,k) &\geq R_1^{-1}(k) \,,\quad k\in \mathcal N_2,\\
\label{cond_sf_f}  \hspace{-4cm} Q_2(h,0) &\geq R_2^{-1}(h) \,,\quad h\in \mathcal N_1  \\
Q_1(N_1+1,k)& > 0\quad k\in \mathcal N_2/ \{N_2\}  \label{sf_g} \\
Q_2(h,N_2+1)& > 0\quad h\in \mathcal N_1/ \{N_1\} \,, \label{sf_h}
\end{align}
\end{subequations}
where
 $L(h,k)=\begin{pmatrix} L_{1}(h,k) & L_{2}(h,k)\end{pmatrix}$.

In this case, a finite-region stabilizing control law has the form \eqref{state_feedback}, with $K(h,k)=L(h,k)Q^{-1}(h,k)$.
\end{theorem}
\begin{proof}
The closed loop connection of system \eqref{2dsys_compact_u} and \eqref{state_feedback} yields
\begin{align} \label{2dsys_compact_sf}
    x^{+}(h,k)& = \left(A(h,k) + B(h,k) K(h,k)\right)x(h,k)\nonumber \\
    &=: A_C(h,k)x(h,k) \,,
\end{align}
where
\begin{equation}\label{conditions_main_sf_bis}
   A_C(h,k) = \begin{pmatrix}
   A_{11}(h,k) + B_{1}(h,k) K_1(h,k) & A_{12}(h,k) + B_{1}(h,k)K_2(h,k) \\
   A_{21} (h,k) + B_{2}(h,k)K_1(h,k) & A_{22}(h,k) + B_{2}(h,k)K_2(h,k)
   \end{pmatrix} \,.
\end{equation}


Applying the statement of Theorem \ref{t1}, the closed loop system is finite-region stable if there exists a symmetric positive definite 2D matrix-valued sequence $P(h,k)=\diag\left(P_1(h,k),P_2(h,k)\right)$, of compatible dimensions, such that
\begin{subequations} \label{conditions_main_sf}
\begin{align}
   A_C^T(h,k)P^{+}(h,k)A_C(h,k)-P(h,k)&<0\,,  \quad (h,k)\in \mathcal N_1\times \mathcal N_2
   \label{classical_a_sf}\\
   P(h,k) &>\Gamma (h,k)\,, \quad (h,k)\in \mathcal N_1\times \mathcal N_2  \label{D2_sf}\\
  {{P}_{1}}(0,k) &\le {{R}_{1}}(k)\,,\quad k\in \mathcal N_2   \label{D3_sf} \\
{{P}_{2}}(h,0)&\le {{R}_{2}}(h)\,, \quad h\in \mathcal N_1  \label{D4_sf} \\
P_1(N_1+1,k)& \geq 0\quad k\in \mathcal N_2/ \{N_2\}  \label{D5_sf} \\
P_2(h,N_2+1)& \geq 0\quad h\in \mathcal N_1/ \{N_1\} \,, \label{D6_sf}
   \end{align}
\end{subequations}
where $A_C(h,k)$  satisfies \eqref{conditions_main_sf_bis}, and $P^{+}(h,k)$ satisfies \eqref{defP+}.


By pre- and post-multiplying condition \eqref{classical_a_sf} by $P^{-1}(h,k)=:Q(h,k)$, we obtain 
\begin{equation}\label{classical_a_sf_1}
     Q(h,k)A_{C}^T(h,k)Q^{+^{-1}}(h,k)A_{C}(h,k)Q(h,k)-Q(h,k)<0 \,,
\end{equation}
which, by using Shur complements arguments, can be rewritten (we omit the argument $(h,k)$ for brevity)
    \begin{align}\label{first_condition_design}
 \begin{pmatrix}
    -Q & QA_{C}^T \\
  A_{C}Q   & -Q^{+}
  \end{pmatrix}
 & = \begin{pmatrix}
    -Q & QA^T + QK^TB^T\\
  AQ+BKQ          & -Q^{+}
  \end{pmatrix} \nonumber\\
 &= \begin{pmatrix}
    -Q & QA^T + L^TB^T\\
  AQ+BL          & -Q^{+}
  \end{pmatrix} <0\,,
    \end{align}
where we have let $K(h,k)Q(h,k)=L(h,k)$; from this last consideration, condition \eqref{cond_sf_a} follows.

By applying the properties of symmetric positive definite matrices, it is readily seen that conditions \eqref{cond_sf_d}--\eqref{cond_sf_f} are equivalent to \eqref{D2_sf}--\eqref{D4_sf}, while conditions \eqref{sf_g}--\eqref{sf_h} imply \eqref{D5_sf}--\eqref{D6_sf}\hfill
\end{proof}

\begin{figure}
  \centering
  \includegraphics[width=8cm]{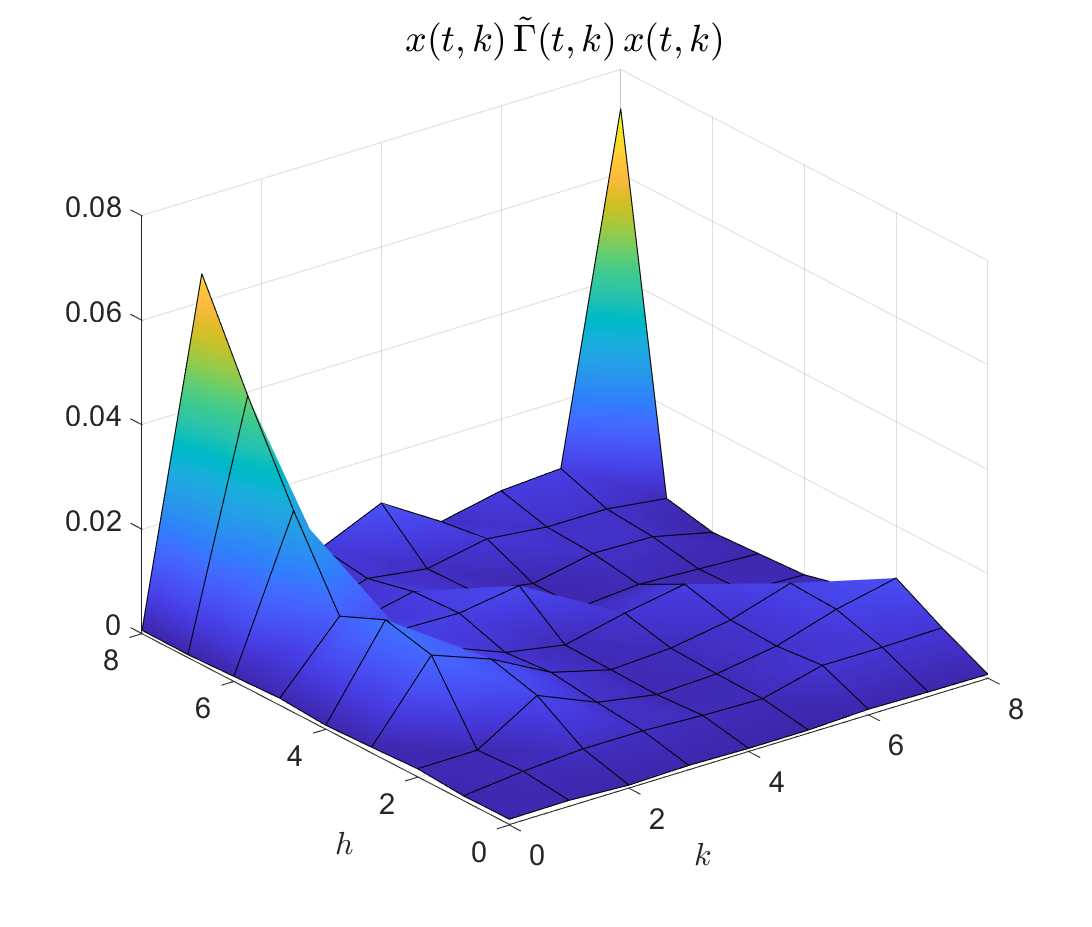}\\
  \caption{The squared weighted norm of the state of the closed
loop system}\label{fig:Ex3_CL_Norm}
\end{figure}

\begin{example}\label{ex:FRS_synthesis}
Let us Consider system \eqref{2dsys_compact_u}, with 
\begin{align*}
    A_{11}(h,k) &= \begin{pmatrix}
  0.6\cos\left(2\left(h+k\right)\right) & 0.5 & 0  \\
   0 & 0.5 & 0.7\sin\left(0.5\left(h+k\right)\right)  \\
   0 & 0 & 0.65  \\
\end{pmatrix}\,,\\ 
A_{12}(h,k) &= \begin{pmatrix}
   0.8\sin\left(0.3\left(h+k\right)\right)  \\
   0  \\
  0  \\
\end{pmatrix}\,,\\
A_{21}(h,k) &= \begin{pmatrix}
   -0.6 & -0.7 & 0.4\sin\left(0.5\left(h+k\right)\right)   \\
\end{pmatrix}\,,\\ 
A_{22}(h,k) &= 0.7\sin\left(2\left(h+k\right)\right),\\
B_{11}(h,k) &= \begin{pmatrix}
    1 + 0.1 h - 0.4 k \\
    0\\
    0.7\\
    \end{pmatrix},\quad B_{21}(h,k)=-1.2\cos\left(0.5\left(h+k\right)\right)\,, 
\end{align*}
and let us set the FRS requirements to $N_1 = N_2 = 8$, $R_1(k)=I, k=0,\dots,N_2$, $R_2(h)=I, h=0,\dots,N_1$, $\Gamma_1(h,k) = \Gamma_2(h,k) = 10^{-2} 1.45^{h+k} I$.

Using the same optimization tools employed in the previous examples, it was not possible to find a solution to the conditions of Theorem \ref{t1}, thus we cannot state that the open-loop system is FRS wrt to $(\mathcal N_1\times \mathcal N_2,R,\Gamma(h,k))$. 

Next, we tackled the FRS synthesis problem, looking for a solution to the conditions of Theorem \ref{t3}, which was found; therefore, Problem \ref{prob_sf} turns out to be feasible for the given system and it is possible to compute a controller gain $K(\cdot,\cdot)$, such that the closed loop system is FRS with respect to $(\mathcal N_1\times \mathcal N_2,R,\Gamma(h,k))$ (see the solution reported in Figure \ref{fig:K_FRS}).
\begin{figure}[tb]
  \centering
  \includegraphics[width=\columnwidth]{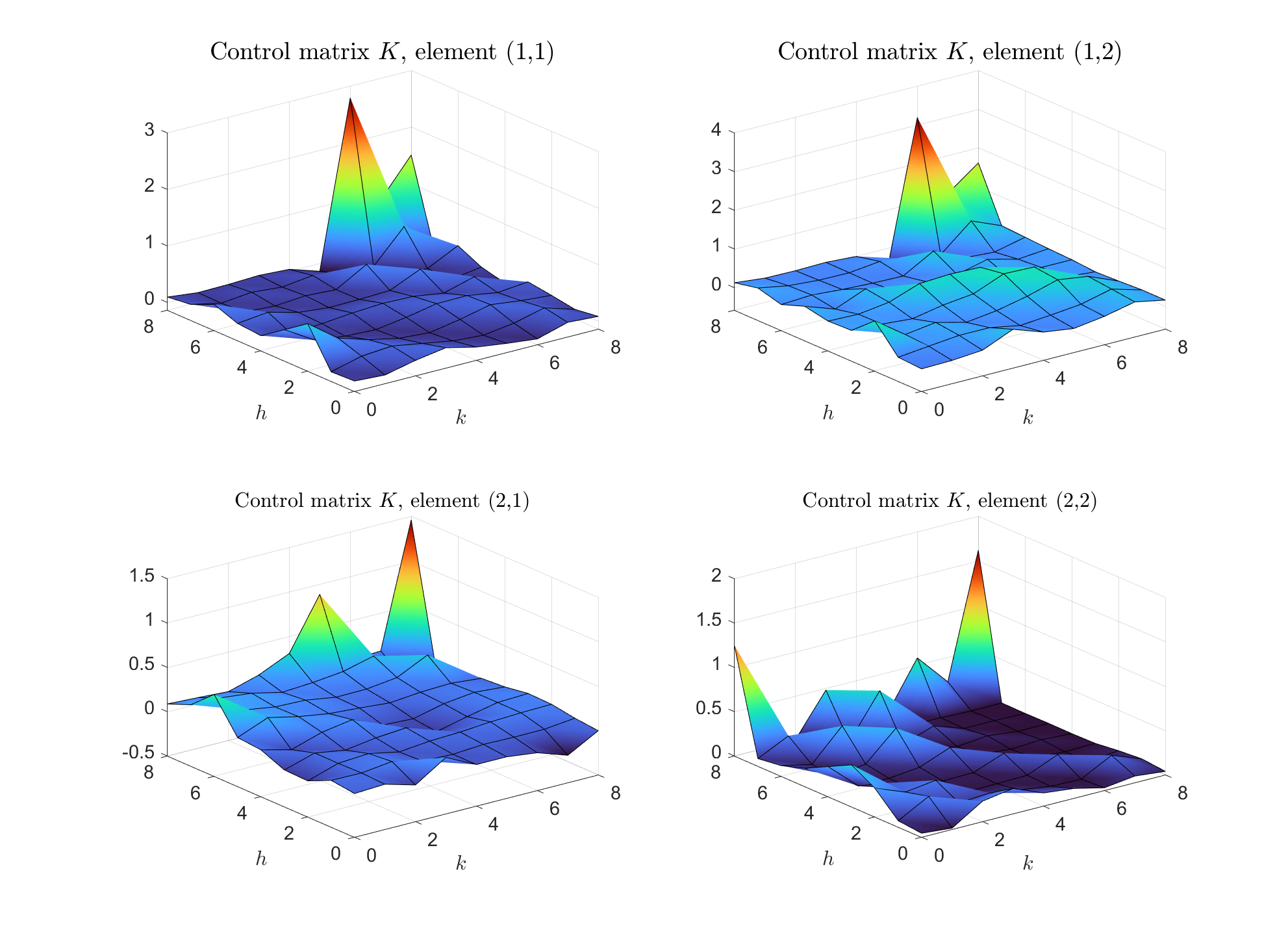}
  \caption{Control function $K(h,k)$ for the closed-loop system designed in Example \ref{ex:FRS_synthesis}.}\label{fig:K_FRS}
\end{figure}
The profile of the squared weighted norm of the state of the closed-loop system, starting from the initial states $x_1(0,k)=\begin{pmatrix}
   0.1361 & 0.1361 & 0.1361   \\
\end{pmatrix}^T$, $x_2(h,0)=0.2357$, $h\in\mathcal N_1$, $k\in \mathcal N_2$, is shown in Fig. \ref{fig:Ex3_CL_Norm}.    
Therefore, the FRS synthesis conditions of Theorem \ref{t3} enabled us to guarantee that the weighted state norm remains below a preassigned threshold over the region of interest.
\end{example}

\FloatBarrier

\section{Conclusions}\label{sec:conclusions}

This paper studies the extension of the FTS methodologies to LTV 2D-systems, namely the FRS approach. The first contribution of the paper is a sufficient condition for FRS, which turns out to be less conservative than the conditions available in the existing literature. Then, an application of the proposed theory to the field of ILC is proposed. To this regard, a procedure to design an ILC law, which guarantees the convergence of the tracking error, within the desired bound, in a finite number of iterations is provided. Finally, starting from the analysis condition, a sufficient condition for the finite-region stabilization of 2D-systems is derived. All the conditions provided in the paper involves the solution of feasibility problems constrained by LMIs, which can be solved via widely available software. Some examples illustrate the effectiveness of the proposed technique.

\section*{References}
\bibliographystyle{IEEEtran}
\bibliography{FTS_old}


\end{document}